\documentclass[aps,prb,twocolumn,superscriptaddress,showpacs,floatfix]{revtex4-2}
\usepackage[T1]{fontenc}
\usepackage[utf8]{inputenc}
\usepackage{amsmath,amssymb,amsfonts,bm}
\usepackage{graphicx}
\usepackage{hyperref}
\usepackage[dvipsnames]{xcolor}
\DeclareRobustCommand{\rev}[1]{\textcolor{black}{#1}}

\begin{document}

\title{Josephson and Spin Currents in Coupled Polariton Condensates}

\author{A.~Kudlis}
\affiliation{Science Institute, University of Iceland, Dunhagi 3, IS-107 Reykjavik, Iceland}

\author{I.~Yu.~Chestnov}
\affiliation{Department of Physics, ITMO University, Saint~Petersburg 197101, Russia}

\author{A.~N.~Osipov}
\affiliation{Department of Physics, ITMO University, Saint~Petersburg 197101, Russia}

\author{A.~V.~Yulin}
\affiliation{Department of Physics, ITMO University, Saint~Petersburg 197101, Russia}

\author{I.~A.~Shelykh}
\affiliation{Science Institute, University of Iceland, Dunhagi 3, IS-107 Reykjavik, Iceland}

\date{\today}

\begin{abstract}
\rev{We analyze particle and spin currents in networks of coupled spinor exciton--polariton condensates arranged as plaquettes and regular polygonal rings. In closed geometries, spin-conserving and TE--TM-induced spin-flip tunnelling combine to generate circulating particle currents, hidden spin counterflows, and bond-dependent spin-current patterns. For the minimal geometries -- an equilateral triangle, and a square plaquette -- we derive analytical expressions for edge-resolved currents from stationary configurations obtained by energy minimization. We then show how particle, in-plane spin, and out-of-plane spin currents partition the parameter plane and provide direct signatures of the equilibrium phases. Finally, we apply the same current-resolved diagnostics to larger rings, where winding numbers and a branch-invariant common-phase coherence metric organize the resulting phase structure.}
\end{abstract}

\maketitle

\section{Introduction}
\allowdisplaybreaks
Strongly correlated many-body systems in lattice geometries provide a powerful framework for engineering and exploring novel phases of matter~\cite{Grenier2002,Georgescu2014}. However, simulation of complex lattice models demands platforms with strong tunable interactions between lattice sites~\cite{Periwal2021}. Conventional systems, such as ultracold atoms~\cite{Bloch2014}, trapped ions~\cite{Monroe2021}, or superconducting circuits~\cite{You2011}, often face limitations in terms of operational tunability and require extreme conditions, such as ultra-low temperatures. In this context, Bose-Einstein condensates of exciton-polaritons (hereafter polaritons) in planar semiconductor microcavities emerge as a highly versatile alternative \cite{Amo2016}. These systems benefit from all-optical control \cite{Alyatkin2020,Dovzhenko2023}, exceptional coherence properties \cite{Topfer2021}, and the ability to realize a wide range of one- and two-dimensional lattice configurations \cite{Kim2011,Milicevic2015,klembt2018exciton,Dusel2020,Pickup2020,Georgakilas2025}.

Polaritons are hybrid light-matter quasiparticles arising from the strong coupling between quantum well excitons and cavity photons. \rev{The combination of their ultra-light effective mass and bosonic character allows condensation at elevated temperatures} \cite{Plumhof2014,Su2020}, while their inherent photonic component facilitates fast non-equilibrium dynamics and direct optical probing \cite{Carusotto2013}. A key advantage of polaritons is their pronounced nonlinearity, originating from the excitonic constituent, which enables strong interparticle interactions. Furthermore, polaritons possess a spin degree of freedom, inherited from the photon polarization and the exciton spin \cite{Kavokin2017_OxfPr}. This spinor character introduces a rich internal structure, allowing polariton lattices to simulate more complex, multi-component physical models. Recent theoretical work has extended the XY lattice model to include this spinor degree of freedom \cite{Kudlis2025}. The application of a magnetic field results in the spin Meissner effect, whose specific manifestations are highly dependent on the underlying lattice geometry~\cite{Chestnov2025}.

The establishment of global phase coherence in a lattice of coupled condensates facilitates the emergence of supercurrents. These persistent currents, driven by phase gradients between sites, are a hallmark of quantum coherence and are central to phenomena like superfluidity and superconductivity. In driven-dissipative systems such as polariton condensates, they additionally serve to compensate for gain-loss imbalances. A key distinction arises in systems possessing a spinor order parameter \cite{Beattie2013}, where the flow of particles can carry a net spin polarization \cite{Pickup2021}. The generation of such spin currents is a subject of great interest, particularly for their potential implications in spintronics.

\rev{In this paper, we theoretically investigate current formation in networks of coupled spinor polariton condensates. We analyze how Josephson-like particle currents, spin transport, and TE--TM-induced torque terms depend on the spin-conserving coupling, spin-flip coupling, Zeeman splitting, and lattice topology. Previous studies have primarily addressed either equilibrium phase structures \cite{Kudlis2025,Chestnov2025} or strongly non-equilibrium condensation mechanisms governed by driven--dissipative processes \cite{Berloff2017}. Here, we focus on conservative equilibrium configurations relevant to high-finesse microcavities \cite{Sun2017,Alnatah2024}. Since thermal equilibrium selects a minimum-energy state, we determine phase diagrams by energy minimization and then interpret the phases through bond-resolved particle and spin currents.}

\rev{The paper is organized as follows. In Sec.~\ref{sec:Model} we introduce the conservative spinor Hamiltonian and define particle and spin currents, including the transport/torque decomposition. In Sec.~\ref{sec:results} we analyze current patterns in a dyad, an equilateral triangle, a square plaquette, and larger polygonal rings, and then discuss the approach to the continuum-ring limit. Section~\ref{sec:Conclusion} summarizes the results.}

\allowdisplaybreaks

\section{Model}\label{sec:Model}

\subsection{Hamiltonian}
We consider a network of $N$ spatially separated exciton-polariton condensates in a magnetic field. Using the second-quantization framework with $\hat{\bm{a}}_j=(\hat a_{j+},\hat a_{j-})^{\mathsf T}$, the total Hamiltonian reads
\begin{subequations}
\begin{align}
\hat H&=\hat H_Z+\hat H_U+\hat H_T,
\\
\hat H_Z&=-\frac{\hbar\Delta_Z}{2}\sum_j \hat{\bm{a}}_j^\dagger\sigma_z\hat{\bm{a}}_j,
\\
\hat H_U&=\hbar U\sum_j\sum_{\sigma=\pm}\hat a_{j\sigma}^\dagger\hat a_{j\sigma}^\dagger\hat a_{j\sigma}\hat a_{j\sigma},
\\
\hat H_T&=\sum_{\langle jl\rangle}\!\Big(\hat{\bm{a}}_j^\dagger T_{jl}\hat{\bm{a}}_l+\hat{\bm{a}}_l^\dagger T_{jl}^\dagger\hat{\bm{a}}_j\Big),
\end{align}
\end{subequations}
where $\Delta_Z$ is the Zeeman splitting, $U$ is the on-site interaction between equal circular components (the opposite-spin interaction is neglected), and $\sigma=\pm$ labels circular polarization. The tunnelling term combines spin-conserving and spin-flip hopping through the $2\times2$ matrix
\begin{align}
T_{jl}=\hbar\Big[J +\delta J\big(\cos\theta_{jl}\sigma_x - \sin\theta_{jl}\sigma_y\big)\Big].
\end{align}
Here $J$ is the spin-conserving hopping amplitude and $\delta J$ is the TE--TM-induced spin-flip amplitude~\cite{Chestnov2025}. The angle $\theta_{jl}$ is twice the polar angle (defined modulo $2\pi$) of the bond connecting sites $j$ and $l$.

In the mean-field limit we replace operators by $c$-numbers, $\hat{\bm{a}}_j\to\bm{\Psi}_j$, where $\bm{\Psi}_j=(\psi_{j,+},\psi_{j,-})^{\mathsf T}$ is the condensate spinor in the circular basis. The classical Hamiltonian $H=\langle\hat H\rangle$ becomes
\begin{subequations}\label{eq:H}
\begin{align}
H&=H_Z+H_U+H_T,
\\
H_Z&=-\frac{\hbar\Delta_Z}{2}\sum_j \bm{\Psi}_j^\dagger\sigma_z\bm{\Psi}_j,
\\
H_U&=\hbar U\sum_j\sum_{\sigma=\pm}\big(\bm{\Psi}_j^\dagger P_\sigma\bm{\Psi}_j\big)^{\!2},
\\
H_T&=\sum_{\langle jl\rangle}\! \Big(\bm{\Psi}_j^\dagger T_{jl}\bm{\Psi}_l+\bm{\Psi}_l^\dagger T_{jl}^\dagger\bm{\Psi}_j\Big),
\end{align}
\end{subequations}
where $P_\pm=(1\pm\sigma_z)/2$. Treating $\bm{\Psi}_j$ and $\bm{\Psi}_j^*$ as canonically conjugate variables gives
\begin{multline} \label{eq:HamiltonEqs}
i\hbar\dot{\bm{\Psi}}_j=\frac{\partial H}{\partial \bm{\Psi}_j^{*}}
= -\frac{\hbar\Delta_Z}{2}\sigma_z\bm{\Psi}_j\\
+2\hbar U\mathrm{diag}\!\big(|\psi_{j,+}|^2,|\psi_{j,-}|^2\big)\bm{\Psi}_j
+\sum_{l\in\langle j\rangle}\! T_{jl}\bm{\Psi}_l,
\end{multline}
\rev{which governs the conservative dynamics. Gain, dissipation, and the dynamical stability of the conservative equilibria are outside the scope of the present paper and are left for a separate study.}

\subsection{Observables: particle and spin currents}

On each directed bond $l\!\to\!j$ we define the particle current $I_{l\to j}$ and the pseudospin current $\bm{J}_{l\to j}$ as
\begin{subequations}\label{eq:current_definitions}
\begin{align}
I_{l\to j}&=\frac{2}{\hbar}\Im\big(\bm{\Psi}_j^\dagger T_{jl}\bm{\Psi}_l\big),\label{eq:part_cur}
\\
\bm{J}_{l\to j}&=\frac{2}{\hbar}\Im\big(\bm{\Psi}_j^\dagger\boldsymbol{\sigma} T_{jl}\bm{\Psi}_l\big)=\bm{J}^{\mathrm{tr}}_{l\to j}+\boldsymbol{\tau}_{l\to j}, \label{eq:spin_current}
\\
\bm{J}^{\mathrm{tr}}_{l\to j}&=\frac{2}{\hbar}\Im\!\Big(\bm{\Psi}_j^\dagger\frac{\{\boldsymbol{\sigma},T_{jl}\}}{2}\bm{\Psi}_l\Big),\label{eq:spin_cur_decomp}
\\
\boldsymbol{\tau}_{l\to j}&=\frac{2}{\hbar}\Im\!\Big(\bm{\Psi}_j^\dagger\frac{[\boldsymbol{\sigma},T_{jl}]}{2}\bm{\Psi}_l\Big),\label{eq:spin_torque}
\end{align}
\end{subequations}
where $\boldsymbol{\sigma}=(\sigma_x,\sigma_y,\sigma_z)^\intercal$ and $\bm{s}_j=\bm{\Psi}_j^\dagger\boldsymbol{\sigma}\bm{\Psi}_j$ is the on-site pseudospin vector. The split into a transport part $\bm{J}^{\mathrm{tr}}_{l\to j}$ and an interfacial torque $\boldsymbol{\tau}_{l\to j}$ is useful because only the former represents spin carried across the edge of the network, while $\boldsymbol{\tau}_{l\to j}$ describes the spin conversion generated by the TE--TM coupling. Under the tunneling direction reversal,
\begin{align*}
I_{l\to j}=-I_{j\to l},\qquad \bm{J}^{\mathrm{tr}}_{l\to j}=-\bm{J}^{\mathrm{tr}}_{j\to l},\qquad \boldsymbol{\tau}_{l\to j}=\boldsymbol{\tau}_{j\to l},
\end{align*}
\rev{so the total spin current $\bm{J}_{l\to j}$ needs not be antisymmetric under bond reversal.}

In the following, we denote the magnitude of the in-plane spin-current component by
\begin{equation*}
J^{\perp}_{l\to j}\equiv \big|\bm{J}^{\perp}_{l\to j}\big|
=\sqrt{(J^x_{l\to j})^2+(J^y_{l\to j})^2}.
\end{equation*}

In what follows we restrict attention to states with equal site occupancies,
$n_j\equiv \bm{\Psi}_j^\dagger\bm{\Psi}_j=n$. We therefore study the currents per particle, $I_{l\to j}/n$ and $\bm{J}_{l\to j}/n$, and omit the explicit factor $n$. The validity of this approximation has been discussed previously in~\cite{Kudlis2025}.

For the analytical stationary states it is convenient to parameterize each site spinor as
\begin{equation}
\psi_{j,\sigma}
=
\sqrt{\frac{n}{2}\bigl(1+\sigma S_{jz}\bigr)}
\,e^{i\left(\Phi_j+\sigma\phi_j/2\right)},
\qquad \sigma=\pm1,
\label{eq:spinor_param}
\end{equation}
\rev{where $S_{jz}=s_{jz}/n=(|\psi_{j,+}|^2-|\psi_{j,-}|^2)/n\in[-1,1]$ is the degree of circular polarization, $\Phi_j=(\varphi_{j,+}+\varphi_{j,-})/2$ is the common phase, and $\phi_j=\varphi_{j,+}-\varphi_{j,-}$ is the relative phase between the two circular components, $\psi_{j,\sigma}=|\psi_{j,\sigma}|e^{i\varphi_{j,\sigma}}$. Equivalently, the densities of circularly polarized components are $n_{j,\pm}=n_j(1\pm S_{jz})/2$. Whenever we write $S_z$ below, we mean the site-independent value $S_{jz}\equiv S_z$ selected by the stationary minimum. The network configuration is then  determined by minimizing the classical Hamiltonian at fixed system parameters $J,\delta J,\Delta_Z,U$ and $n$: $\Delta_Z$ biases $S_z$, $U$ enters the nonlinear precession, and $J$ together with $\delta J$ fixes the phase locking of the condensates.}

Equation~\eqref{eq:HamiltonEqs} yields the local continuity equations
\begin{subequations}
\begin{align}
\dot n_j&=\sum_{l\in\langle j\rangle} I_{l\to j}, \label{eq:PartContinuity}\\
\dot{\bm{s}}_j&=\sum_{l\in\langle j\rangle}\bm{J}_{l\to j}
- \big(\Delta_Z-2U s_{jz}\big) \hat{\bm{z}}\times\bm{s}_j, \label{eq:SpinContinuity}
\end{align}
\end{subequations}
where Eq.~\eqref{eq:PartContinuity} constitutes conservation of the total number of particles. By contrast, pseudospin is not separately conserved: the last term in Eq.~\eqref{eq:SpinContinuity} is the local Zeeman- and interaction-induced (self-induced Larmor) precession.

\rev{We are primarily interested in the effects of the magnetic field and TE--TM splitting, so it is useful to exploit the following symmetries to restrict the parameter plane.} First, the sign change $\delta J \to -\delta J$ is generated by the unitary transformation $\bm{\Psi}_j \to \sigma_z \bm{\Psi}_j$, which corresponds to a rotation by $\pi$ around the $z$-axis in the Stokes vector space. Indeed, this transformation leaves $\hat H_Z$ and $\hat H_U$ invariant, while for the tunnelling term $\sigma_z T_{jl}(\delta J) \sigma_z = T_{jl}(-\delta J)$. Consequently, the energy spectra for $\delta J$ and $-\delta J$ are identical. In terms of physical observables, this transformation preserves the particle current and the $z$-component of the spin current (circular polarization), while reversing the $x$ and $y$ components. 
Second, the energy spectrum is invariant under the combined operation of time reversal $\bm{\Psi}_j\to\sigma_x\bm{\Psi}_j^{*}$ and sign reversal of the magnetic field, $\Delta_Z\to-\Delta_Z$.
Under this map, $I_{l\to j}$, $J^x_{l\to j}$, and $J^y_{l\to j}$ reverse sign and $J^z_{l\to j}$ remains unchanged. We therefore address only the domain $\delta J\ge0$ and $\Delta_Z\ge0$. \rev{Configurations at other signs of $\delta J$ or $\Delta_Z$ can be reconstructed using the symmetry transformations discussed above.}

\section{Results for prototypical geometries}\label{sec:results}

\rev{In equilibrium, the stationary configuration minimizes the total energy. For the dyad, triangle, and square, the ground-state configurations were found in Refs.~\cite{Kudlis2025,Chestnov2025}. Below we introduce them in terms of the variables $(\Phi_j,\phi_j,S_{jz})$ given by Eq.~\eqref{eq:spinor_param} and focus on the resulting bond currents. For larger systems, a complete phase-by-phase parametrization becomes unwieldy; we therefore obtain the stationary minima numerically and classify them using particle currents, spin currents, polarization-resolved winding numbers, and a branch-invariant common-phase coherence metric.}

\subsection{Currents in a spinor dimer (dyad)}
\label{subsec:curr-dimer}

Two coupled condensates in the ground state are phase-locked with $\Phi_2-\Phi_1=0$ for $J<0$ and $\Phi_2-\Phi_1=\pi$ for $J>0$. The density continuity \eqref{eq:PartContinuity} then enforces absence of the net particle flow:
\begin{equation*}
I_{1\to2}=I_{2\to1}=0.
\end{equation*}
To compensate for the external magnetic field, the condensates become elliptically polarized, $S_{1z}=S_{2z}\neq0$. It involves a self-induced rotation of the pseudospin about the $z$ axis, which must be balanced by the in-plane spin currents, according to Eq.~\eqref{eq:SpinContinuity}.  
Then, from the definition \eqref{eq:spin_current}, one obtains
\begin{subequations}
\begin{align}
&\{J^x_{1\to2},J^y_{1\to2}\}
=-2\chi \delta J S_z\{\sin\theta_{12},\cos\theta_{12}\},\\
&J^z_{1\to2}=0,
\end{align}   
\end{subequations}
where $\chi\equiv\cos(\Phi_2-\Phi_1)=\pm1$ and the bond-reversal symmetry $\bm{J}_{1\to2}=\bm{J}_{2\to1}$ holds. The latter implies that the spin current contains only the torque part and exhibits no spin transport.

\rev{Networks of several coupled condensates are substantially more complex than a dyad and can support nontrivial current patterns. We focus on closed-loop geometries. For a stationary equal-density state on a regular polygon, Eq.~\eqref{eq:PartContinuity} enforces the same net particle current on every directed edge of the ring. Spin currents are more intricate: the transported spin and the TE--TM-induced interfacial torque can vary from bond to bond even when the particle circulation is uniform. In the next sections we analyze how these persistent particle and spin currents appear in polygonal geometries.}

\subsection{Currents in an equilateral triangle}
\label{subsec:curr-triangle}

\begin{figure}
    \centering
    \includegraphics[width=0.9\linewidth]{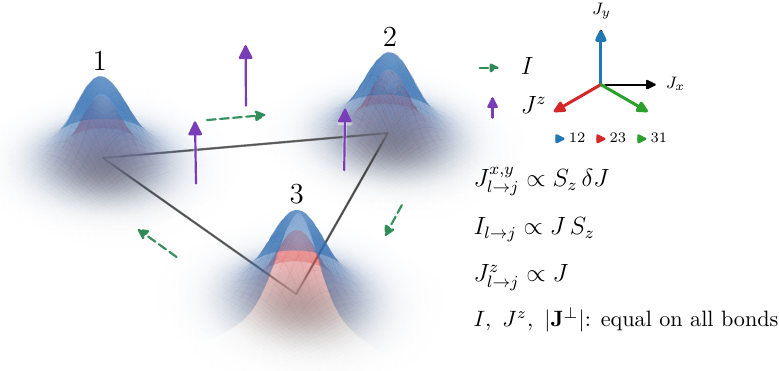}
\caption{\rev{Schematic of the hidden-vortex triangular state (phase~III). The two circular components ($\sigma_{+}$ and $\sigma_{-}$, blue/red lobes) wind with opposite chiralities. At finite circular polarization this produces
a uniform particle current $I_{l\to j}$ along the directed ring, while the out-of-plane spin current $J^z_{l\to j}=\sqrt{3}J$ is uniform on all three bonds. The inset gives the in-plane Stokes-space spin-current vectors $\bm{J}^{\perp}_{l\to j}$; their directions rotate from bond to bond, but their magnitude is the same, $J^{\perp}_{l\to j}=2|\delta J S_z|$. Green dashed arrows denote particle flow and purple arrows denote the $z$-spin current; reversing the circulation convention or the sign of $J$ reverses the corresponding arrow directions.}}
\label{fig:triangle_spin_currents_3d}
\end{figure}

The simplest closed geometry is the equilateral triangle. We label the sites clockwise as $1,2,3$ and choose the bond angles:
\begin{equation*}
\theta_{12}=\frac{2\pi}{3},\qquad \theta_{23}=\frac{4\pi}{3},\qquad \theta_{31}=0.
\end{equation*}
The stationary minima of $H$ given by Eq.~\eqref{eq:H} found in Ref.~\cite{Chestnov2025} comprise an asymmetric state (phase~I), a semi-vortex state (phase~II), and a hidden-vortex state (phase~III).

\paragraph*{Hidden-vortex state (phase III).}
In this state the two circular components wind with opposite chiralities while $S_{jz}\equiv S_z$ is uniform. This configuration supports a field-induced net particle circulation, which vanishes at $S_z=0$:
\begin{equation}
I_{\rm III}=\sqrt{3}JS_z.
\end{equation}
The total particle circulation is therefore $3\sqrt{3}JS_z$. At weak fields, the elliptic polarization degree reads \cite{Chestnov2025}:
\begin{equation*}
S_z \approx \frac{\Delta_Z}{2\left( 2 |\delta J| + Un \right)},
\end{equation*}
so the particle current is linear in $\Delta_Z$ to leading order. The out-of-plane spin current is
\begin{equation}
   J^{z}_{1\to 2}=J^{z}_{2\to 3}=J^{z}_{3\to 1} =  \sqrt{3}J,
\end{equation}
independent of $S_z$ for equal occupancies. For the ferromagnetic case, $J<0$, which is shown in the phase maps below, this corresponds to the counterclockwise circulation, $J^z_{1\to2} <0$. More generally, for a regular $m$-edge polygon with opposite chiralities in the two circular components, the bond current is $J^z = 2J\sin(2\pi/m)$, so the total $z$-spin circulation tends to $4\pi J$ in the continuous-ring limit. The in-plane components are
\begin{subequations}
\begin{align}
    \{ J^{x}_{1\to 2},J^{x}_{2\to 3},J^{x}_{3\to 1} \} &= \{0,-\sqrt{3},\sqrt{3}\}  \delta J S_z, \\
    \{ J^{y}_{1\to 2},J^{y}_{2\to 3},J^{y}_{3\to 1} \} &= \{-2,1, 1\}   \delta J S_z,
\end{align}
\end{subequations}
with the bond-independent magnitude $J^{\perp}_{l\to j}=2|\delta J S_z|$.

\begin{figure}
    \centering
\includegraphics[width=1.0\linewidth]{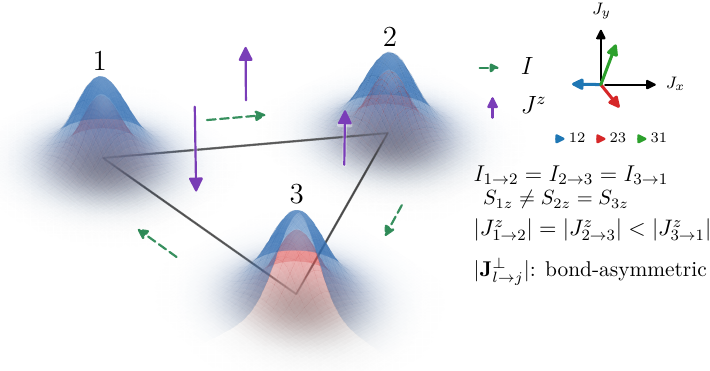}
\caption{\rev{Schematic of the asymmetric triangular state (phase~I). The condensate on site~1 has a different circular polarization from sites~2 and~3, and the equilibrium phases generate a uniform particle current around the loop but a bond-asymmetric spin-current texture. Green dashed arrows denote $I_{l\to j}$, purple arrows indicate $J^z_{l\to j}$, and the inset shows the in-plane Stokes-space vectors $\bm{J}^{\perp}_{l\to j}$. The two bonds related by the residual symmetry have equal $|J^z|$, whereas the $3\!\to\!1$ bond carries a different out-of-plane spin current; the in-plane directions also change from bond to bond.}}
    \label{fig:tri_phase1}
\end{figure}
\begin{figure}
    \centering
    \includegraphics[width=1\linewidth]{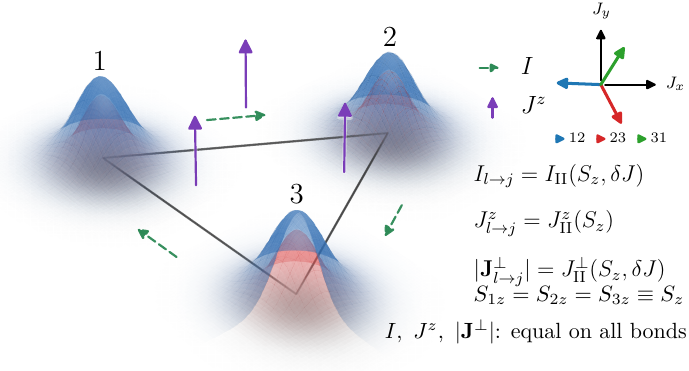}
\caption{\rev{Schematic of the semi-vortex triangular state (phase~II). Only one circular component carries a $2\pi/3$ phase winding, so the particle current is uniform on all three directed bonds, while the $z$-spin current reflects the circulation imbalance between the circular components. Green dashed arrows show $I_{l\to j}$, purple arrows show $J^z_{l\to j}$, and the inset gives $\bm{J}^{\perp}_{l\to j}$ in Stokes space. In contrast to phase~I, the particle current, $J^z_{l\to j}$, and the magnitude of $\bm{J}^{\perp}_{l\to j}$ are bond independent.}}
    \label{fig:triangular_phase_2}
\end{figure}

\paragraph*{Asymmetric state (phase I).}
\rev{Phase~I exists only for the ferromagnetic sign of the spin-conserving coupling, $J<0$. Its stationary spinors can be parametrized as}
\begin{align*}
(\Phi_1,\Phi_2,\Phi_3)&=(0,\delta\alpha,-\delta\alpha),\\
(\phi_1,\phi_2,\phi_3)&=(2\phi_0,2\phi_0-\delta\phi,2\phi_0+\delta\phi),
\end{align*}
with
\begin{equation*}
(S_{1z}, S_{2z}, S_{3z} ) = (S_z+\delta S_z,S_z,S_z),
\end{equation*}
where $\phi_0$ is an overall relative-phase offset and $\delta S_z$, $\delta\alpha$, and $\delta\phi$ are fixed by energy minimization. The uniform particle current is governed by:
\begin{align}
&I_{\mathrm{I}}
=2\Big[J\left(\cos(\delta\phi)\sin(2\delta\alpha)-S_z\cos(2\delta\alpha)\sin(\delta\phi)\right)\nonumber\\
&\qquad\qquad\qquad\qquad\qquad- \left|\delta J\right|\sin(2\delta\alpha)\sqrt{1-S_z^2}
\Big].
\label{eq:FM1_total_current}
\end{align}
Note that the current vanishes at zero magnetic field, $I_{\mathrm{I}}=0$, when the symmetry of the phase pattern is restored,
$S_z=\delta S_z=\delta\alpha=\delta\phi=0$.

\rev{Closed expressions for the spin current in phase~I are lengthy and not very informative, so we summarize only the robust structure. According to the continuity condition \eqref{eq:SpinContinuity}, the spin current balances the magnetic-field- and interaction-induced precession of the pseudospin and therefore has all three non-vanishing projections. The spin currents also inherit the asymmetry of the phase structure: the current on the edge $3\!\to\!1$ differs from those on $1\!\to\!2$ and $2\!\to\!3$, as summarized in Fig.~\ref{fig:tri_phase1}.}

\begin{figure}
    \centering
    \includegraphics[width=1\linewidth]{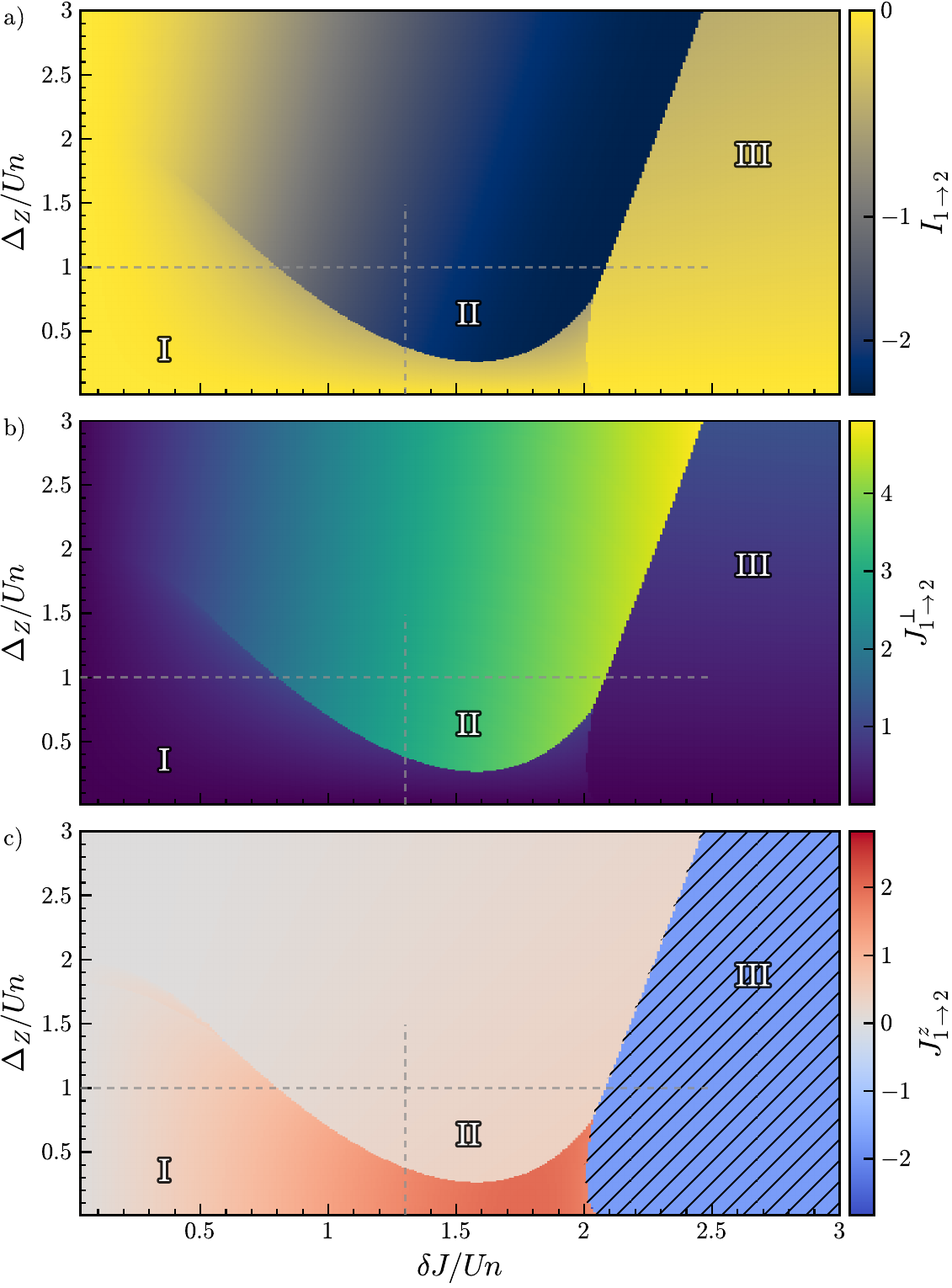}
\caption{\rev{Ground-state current maps for an equilateral triangle at fixed $J/Un=-1$ on the parameter plane $(\delta J/Un,\Delta_Z/Un)$. All values are currents per particle along the  $1\!\to\!2$ edge: (a) particle current $I_{1\to2}$, (b) in-plane spin-current magnitude $J^{\perp}_{1\to2}$, and (c) out-of-plane spin-current component $J^z_{1\to2}$. Roman numerals indicate the energy-minimizing phases introduced in the text (according to the previous author's paper~\cite{Chestnov2025}). The hatched region in panel~(c) is the hidden-vortex phase~III, where $J^z_{1\to2}=\sqrt{3}J$ is fixed by the phase winding and is therefore negative for $J/Un=-1$. Gray dashed guides mark the one-dimensional cuts plotted in Fig.~\ref{fig:triangle_linecuts_currents}; negative $I_{1\to2}$ means flow opposite to the chosen bond orientation.}} \label{fig:triangle_heatmap_I12}
\end{figure}

\begin{figure}
    \centering
\includegraphics[width=1\linewidth]{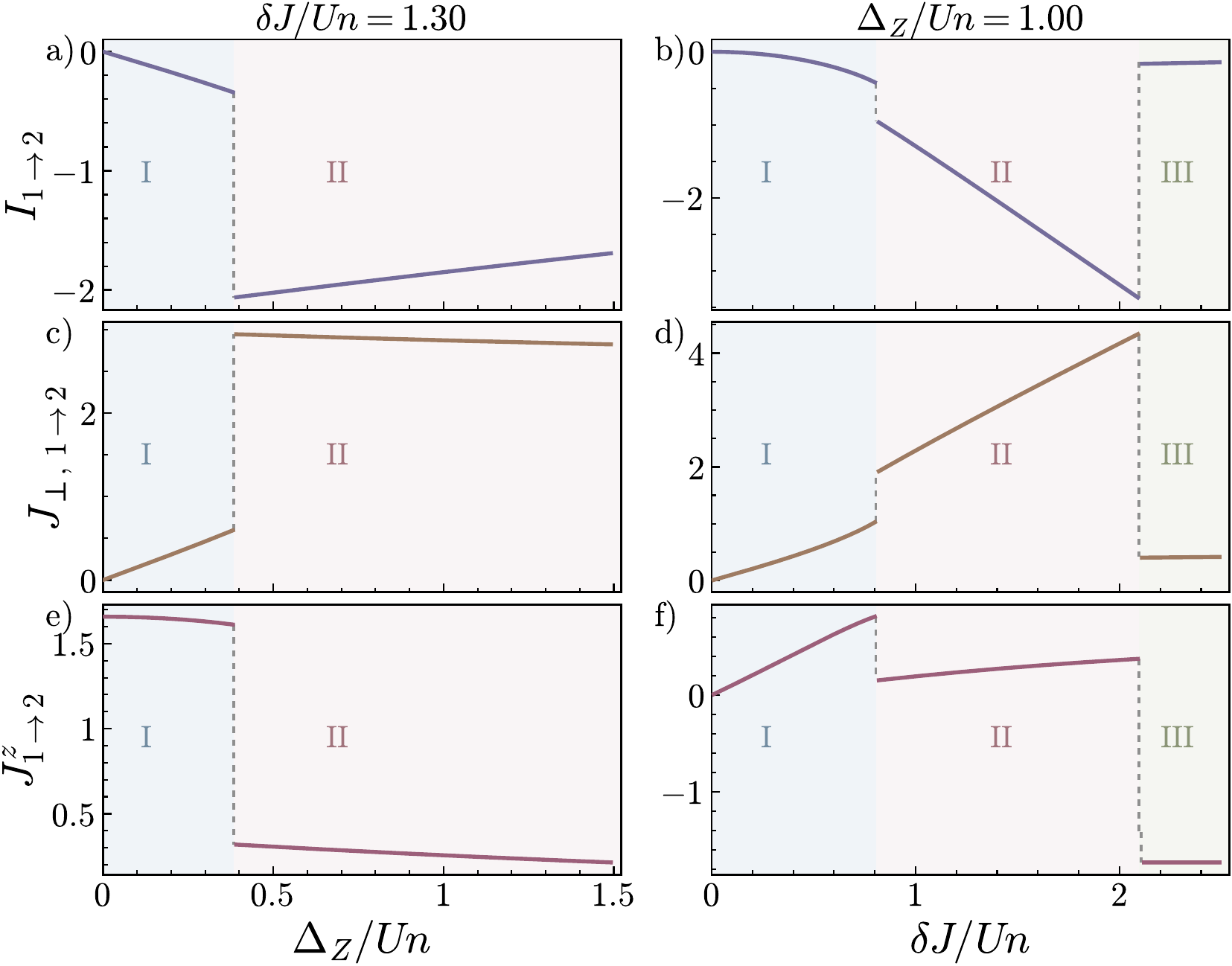}
\caption{\rev{One-dimensional cuts through the triangular maps of Fig.~\ref{fig:triangle_heatmap_I12} at fixed $J/Un=-1$. The left column scans $\Delta_Z/Un$ at fixed $\delta J/Un=1.30$, and the right column scans $\delta J/Un$ at fixed $\Delta_Z/Un=1.00$. Rows show, from top to bottom, $I_{1\to2}$, $J^{\perp}_{1\to2}$, and $J^z_{1\to2}$ on the directed bond $1\!\to\!2$. Shaded backgrounds label the phase that minimizes the energy along each cut, while gray vertical dashed lines mark the corresponding branch-switching points. Discontinuities or sign changes occur at these phase switches; within a fixed branch the current diagnostics vary smoothly.}}
\label{fig:triangle_linecuts_currents}
\end{figure}

\paragraph*{Semi-vortex state (phase II).}
In the ferromagnetic regime $J<0$, \rev{phase~II has a uniform degree of circular polarization, $S_{jz}\equiv S_z$, and a $2\pi/3$ phase winding in the $\sigma_{-}$ component only, while $\sigma_+$ has no phase circulation.} In the positive-$\delta J$ scan considered below, the energy-minimizing semi-vortex phase carries a uniform circulation. For this branch, the particle current at each bond reads
\begin{equation}
  I_{\mathrm{II}}
  =\frac{\sqrt{3}}{2}\Big[
    J\big(1-|S_z|\big)
    -2 |\delta J|\sqrt{1-S_z^2}
  \Big].
  \label{eq:FM2_particle_current}
\end{equation}

The out-of-plane spin current is also uniform,
\begin{equation}
  J^z_{\mathrm{II}}
  =\frac{\sqrt{3}}{2} J\big(|S_z|-1\big)
  \label{eq:FM2_Jz}
\end{equation}
and corresponds to the transport current according to the definition \eqref{eq:current_definitions}, i.e.  $J^{z,\mathrm{tr}}_{l\to j}=J^z_{l\to j}$ with $\tau^z_{l\to j}=0$.

Note that the TE--TM splitting affects $J^z$ only indirectly through the selected value of $S_z$. In particular, $J^z_{\mathrm{II}}$ remains finite even at $S_z=0$ reflecting the spin-conserving current carried by the winding circular component and vanishes at $|S_z|\to1$.

The in-plane spin current $J^{\perp}_{l\to j}$ has a bond-independent magnitude governed by
\begin{align}
  \bigl(J^{\perp}_{l\to j}\bigr)^2
  &=\delta J^2\big(3+S_z^2\big)
  +3J\delta J\big(S_z-1\big)\sqrt{1-S_z^2}
  \nonumber\\
  &\quad +3J^2\big(1-S_z^2\big).
  \label{eq:FM2_Jperp_sq}
\end{align}
Thus $\bigl(J^{\perp}_{l\to j}\bigr)^2\to4\delta J^2$ as $|S_z|\to1$.

\paragraph*{Numerical maps and analytic line cuts.}
\rev{A current-resolved phase diagram for a triangular network, calculated numerically at fixed $J/Un=-1$ in the $(\delta J/Un,\Delta_Z/Un)$ plane for $\delta J>0$, is summarized in Figs.~\ref{fig:triangle_heatmap_I12} and~\ref{fig:triangle_linecuts_currents}. Figure~\ref{fig:triangle_heatmap_I12} shows representative characteristics -- $I_{1\to2}$, $J^{\perp}_{1\to2}$, and $J^z_{1\to2}$ -- on the edge $1\!\to\!2$ between the first and the second condensate. These observables distinguish the three phases, while the bond-asymmetric spin texture of phase~I should be considered together with the full pattern shown in Fig.~\ref{fig:tri_phase1}. Figure~\ref{fig:triangle_linecuts_currents} shows one-dimensional cuts of the same maps. Sharp changes or sign reversals of the currents indicate phase boundaries; along each cut, the observables jump only when the energy-minimizing branch changes.}

\begin{figure}[b]
    \centering
    \includegraphics[width=1\linewidth]{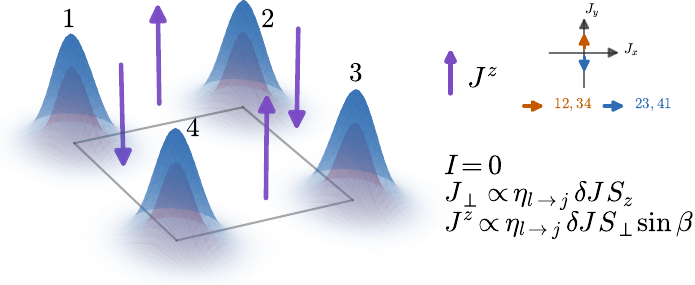}
\caption{\rev{Square-plaquette phase~I: bond-staggered spin current without particle transport. The blue/red lobes show the two circular components on sites $1$--$4$; no green bond arrow is drawn because $I_{l\to j}=0$ on every edge. Purple arrows indicate the $z$-spin current, whose sign alternates between the two edge pairs and whose amplitude depends on the free polarization angle $\beta$ through $J^z\propto\eta_{l\to j}\delta J S_\perp\sin\beta$. The inset summarizes the in-plane spin current: for the chosen square orientation it is purely a $J_y$ Stokes-space component with opposite signs on the two bond pairs.}}\label{fig:square_phaseI_3d}
\end{figure}

\subsection{Currents in a square plaquette}
\label{subsec:curr-square}

We now consider a square plaquette of four condensates labeled $j=1,2,3,4$ and connected cyclically as $1\!\to\!2\!\to\!3\!\to\!4\!\to\!1$. The equilibrium phases were classified in Refs.~\cite{Kudlis2025,Chestnov2025}; here we only discuss the resulting currents. In the convention of Sec.~\ref{sec:Model}, we set
\begin{equation}
\theta_{12}=\theta_{34}=\pi,
\qquad
\theta_{23}=\theta_{41}=0,
\end{equation}
so the spin-flip contribution alternates in sign between adjacent edges.  It is convenient to define
\begin{align}
\eta_{l\to j}&=
\begin{cases}
+1,& (l\to j)\in\{1\to2,\,3\to4\},\\
-1,& (l\to j)\in\{2\to3,\,4\to1\},
\end{cases}
\label{eq:square_eta_def}
\end{align}
which encodes this alternation. For all square phases considered below, energy minimization gives a uniform out-of-plane spin component, $S_{jz}\equiv S_z$.

\begin{figure}[t!]
    \centering
\includegraphics[width=\linewidth]{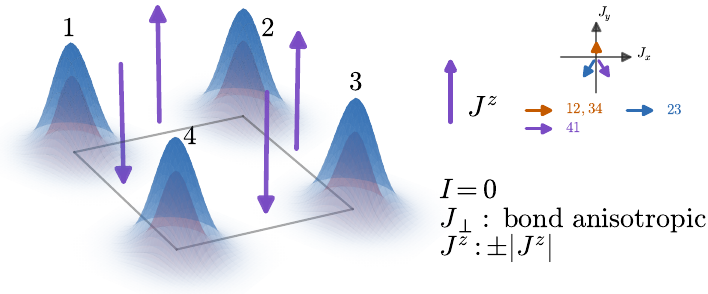}
\caption{\rev{Square-plaquette phase~II: bond-asymmetric spin transport with zero particle current. Stationarity fixes the phase parameter $\alpha$ so that $I_{l\to j}=0$ on all four bonds. The out-of-plane spin current $J^z_{l\to j}$ is finite and alternates in sign between opposite directed bonds, as indicated by the purple arrows. The inset shows the in-plane spin current: bonds $(1\!\to\!2,\,3\!\to\!4)$ carry a purely $J_y$ contribution, whereas bonds $(2\!\to\!3)$ and $(4\!\to\!1)$ have equal $J_y$ and opposite $J_x$. The symmetry-related state $\alpha\to-\alpha$ reverses $J_x$ and $J^z$ but leaves $J_y$ unchanged.}}\label{fig:square_phaseII_3d}
\end{figure}

\begin{figure}[t]
\centering
\includegraphics[width=\linewidth]{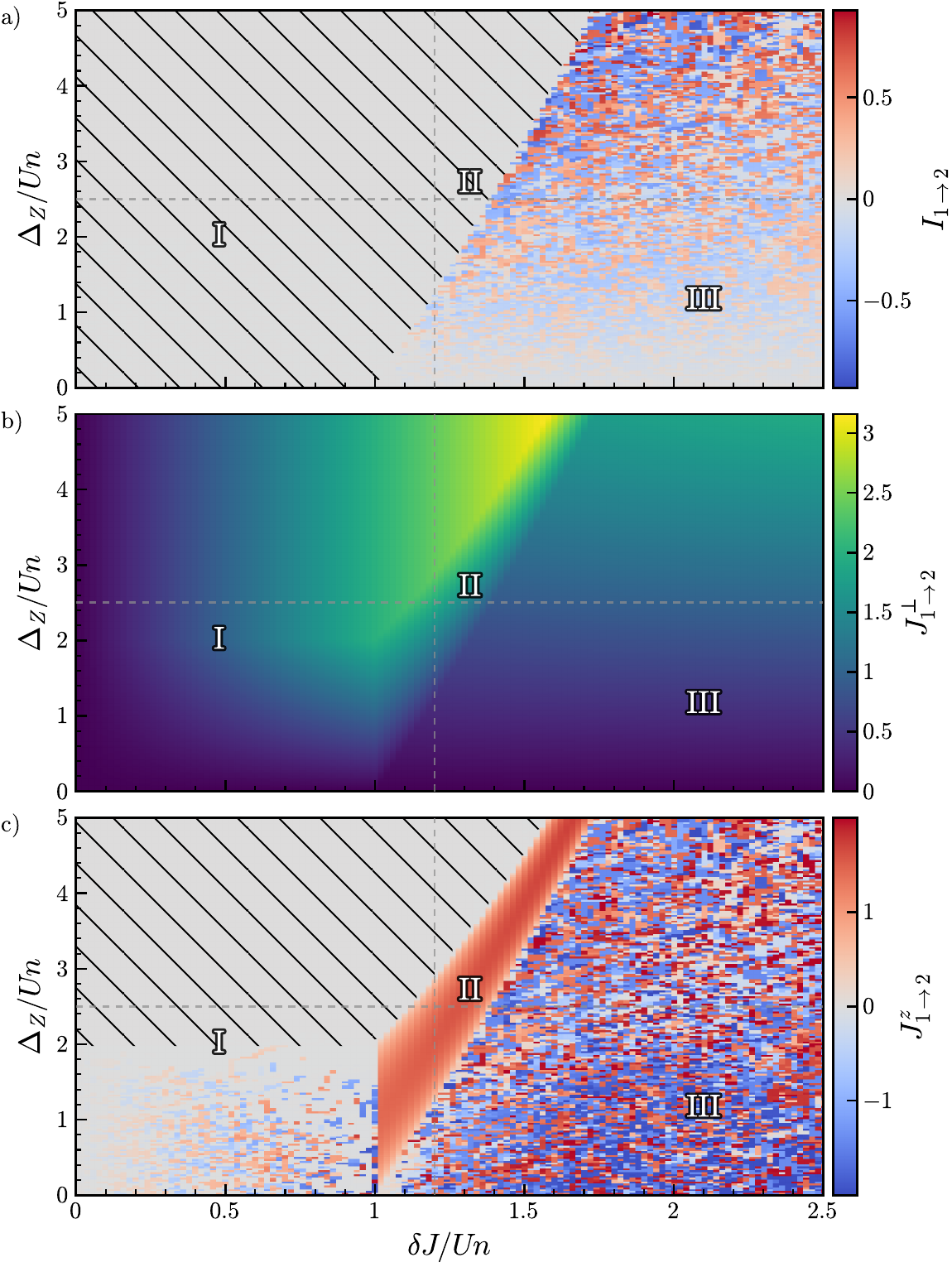}
\caption{\rev{Ground-state current maps for a square plaquette at fixed $J/Un=-1$ in the parameter plane $(\delta J/Un,\Delta_Z/Un)$. All values are currents per particle on the directed bond $1\!\to\!2$: (a) particle current $I_{1\to2}$, (b) in-plane spin-current magnitude $J^{\perp}_{1\to2}$, and (c) out-of-plane spin-current component $J^z_{1\to2}$. In panel~(a), hatching marks the phases where the particle current vanishes identically on every bond (Phases~I and II). In panel~(c), the hatched part of Phase~I marks the fully polarized region, $S_z=1$, where $S_\perp=0$ and hence $J^z_{1\to2}=0$. Gray dashed guides indicate the one-dimensional cuts in Fig.~\ref{fig:square_linecuts_currents}. Fine-scale speckling in the $\beta$-dependent observables reflects the continuous free-angle degeneracy; the in-plane magnitude $J^{\perp}_{1\to2}$ is insensitive to this representative choice.}}
\label{fig:square_heatmaps_three_panel}
\end{figure}

\paragraph*{Bond-staggered torque-induced spin current without particle current (phase I).}
In Phase~I, the total and relative phases are spatially uniform:
\begin{equation*}
\Phi_1=\Phi_2=\Phi_3=\Phi_4,
\qquad
\phi_1=\phi_2=\phi_3=\phi_4=\beta,
\end{equation*}
with the total energy being independent of the angle $\beta$, which, however, defines the orientation of the polarization ellipses \cite{Kudlis2025,Chestnov2025}. Because of this degeneracy, the $\beta$-dependent observables are not uniquely fixed. Substituting the stationary spinors into Eqs.~\eqref{eq:part_cur}--\eqref{eq:spin_torque} yields
\begin{subequations}
\begin{align}
I_{l\to j}&=0, \\
J^x_{l\to j}&=0, \\
J^y_{l\to j}&=2\,\eta_{l\to j}\,\delta J\,S_z, \\
J^z_{l\to j}&=2\,\eta_{l\to j}\,\delta J\,S_\perp\,\sin\beta,
\end{align}
\label{eq:square_phaseI_currents}
\end{subequations}
where $S_\perp \equiv \sqrt{1-S_z^2}$. 
Thus phase~I has no particle transport, while the spin current is staggered from bond to bond through $\eta_{l\to j}$. The fact that $J^x=0$ but $J^y\neq0$ is not fundamental; it follows from the assumed orientation of square edges $\theta_{jl}\in\{0,\pi\}$ and would be exchanged by an in-plane rotation of the pseudospin basis.

\begin{figure}[t]
\centering
\includegraphics[width=\linewidth]{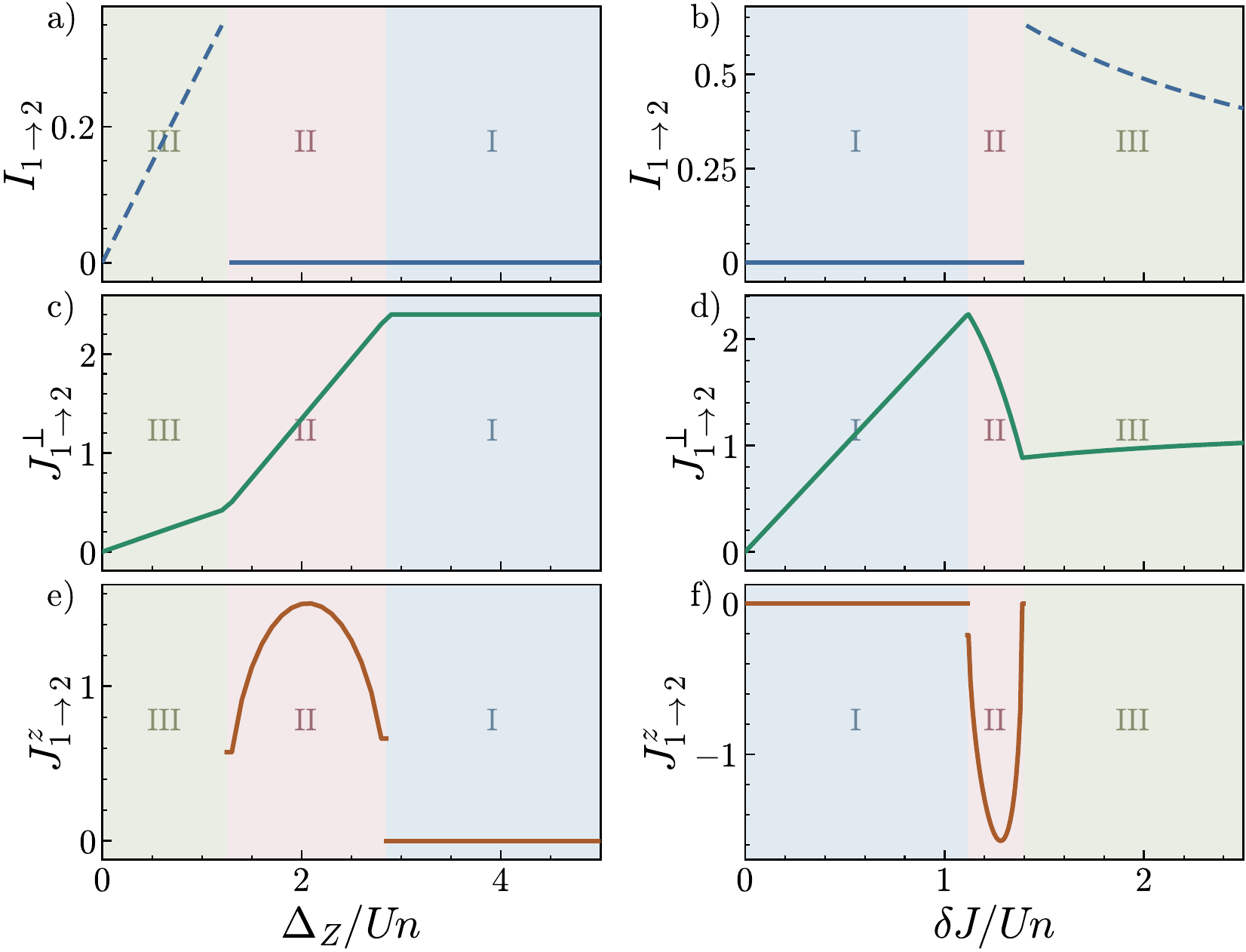}
\caption{\rev{One-dimensional cuts through the square-plaquette maps of Fig.~\ref{fig:square_heatmaps_three_panel} at fixed $J/Un=-1$. The left column scans $\Delta_Z/Un$ at fixed $\delta J/Un=1.2$, and the right column scans $\delta J/Un$ at fixed $\Delta_Z/Un=2.5$. Rows show $I_{1\to2}$, $J^{\perp}_{1\to2}$, and $J^z_{1\to2}$ on the directed bond $1\!\to\!2$. Shaded backgrounds indicate the equilibrium phases I--III, and gray vertical dashed markers denote the analytic phase boundaries. In the $\beta$-degenerate phases, branch-dependent observables are omitted unless a representative is fixed; in particular, $J^z_{1\to2}$ is omitted in Phases~I and III, while the dotted Phase~III segment of $I_{1\to2}$ shows the representative choice $\sin\beta=1$.}}\label{fig:square_linecuts_currents}
\end{figure}

\paragraph*{Bond-asymmetric spin transport with vanishing particle current (phase II).}
It is characterized by
\begin{equation*}
\Phi_1=\Phi_2=+\alpha,
\qquad
\Phi_3=\Phi_4=-\alpha,
\end{equation*}
and
\begin{equation*}
\phi_1=\phi_2=2\alpha-\pi,
\qquad
\phi_3=\phi_4=-2\alpha+\pi,
\end{equation*}
with uniform $S_{jz}\equiv S_z$ and $\alpha$ governed by
\begin{equation}
J(1+S_z)\sin(4\alpha)
=
2\,\delta J\,\sqrt{1-S_z^2}\,\sin(2\alpha).
\label{eq:square_phaseII_stationarity}
\end{equation}
The net particle current vanishes in this phase,
\begin{equation}
I_{1\to2}=I_{2\to3}=I_{3\to4}=I_{4\to1}=0,
\label{eq:square_phaseII_I}
\end{equation}
while the spin currents take the compact form
\begin{subequations}
\begin{align}
J^x_{1\to2}&=J^x_{3\to4}=0,
\nonumber\\
J^x_{2\to3}&=-J^x_{4\to1}
=
\frac{\delta J^2-J^2}{J \delta J}\,J^z_{\mathrm{II}},
\\
J^y_{1\to2}&=J^y_{3\to4}=2\,\delta J\,S_z,
\nonumber\\
J^y_{2\to3}&=J^y_{4\to1}
=
2\,\delta J
\left[
\frac{\delta J^2}{J^2}(1-S_z)-1
\right],
\\
J^z_{1\to2}&=J^z_{2\to3}
=
-\,J^z_{3\to4}
=
-\,J^z_{4\to1}
\equiv J^z_{\mathrm{II}},
\end{align}
\label{eq:square_phaseII_currents}
\end{subequations}
where
\begin{equation}
\bigl(J^z_{\mathrm{II}}\bigr)^2
=
4\,\delta J^2\,(1-S_z)
\left[
(1+S_z)-\frac{\delta J^2}{J^2}(1-S_z)
\right].
\label{eq:square_phaseII_Jz}
\end{equation}
\rev{The phase~II configuration is degenerate with respect to the transformation $\alpha\to-\alpha$.} For this state, the $J^y$ component remains unchanged, while $J^x$ and $J^z$ reverse sign. Phase~II thus carries a robust edge-dependent spin current while keeping $I_{l\to j}=0$ on every bond.

\paragraph*{Uniform circulating state with a free-angle manifold (phase III).}
\rev{Phase~III contains a free-angle manifold of representatives, including states with genuine circulation around the plaquette. The total phases are}
\begin{equation*}
\Phi_1=\Phi_2=\Phi_3=\Phi_4,
\end{equation*}
with alternating relative phases read:
\begin{equation*}
(\phi_1,\phi_2,\phi_3,\phi_4)=(\beta-\pi,-\beta+\pi,\beta+\pi,-\beta-\pi).
\end{equation*}
As in phase~I, the energy is independent of $\beta$, so the $\beta$-dependent currents are not uniquely fixed inside the degenerate manifold. The bond currents are
\begin{subequations}
\begin{align}
&I_{1\to2}=I_{2\to3}=I_{3\to4}=I_{4\to1}=-2J S_z\sin\beta,\\
&J^z_{1\to2}=J^z_{2\to3}=J^z_{3\to4}=J^z_{4\to1}=-2J\sin\beta,\\
&J^x_{l\to j}=2\eta_{l\to j}\delta J S_z\sin\beta,\\
&J^y_{1\to2}=J^y_{2\to3}=J^y_{3\to4}=J^y_{4\to1}=-2\delta J S_z\cos\beta.
\end{align}
\label{eq:square_phaseIII_currents}
\end{subequations}
\noindent \rev{Thus any representative with $\sin\beta\neq0$ carries a uniform $z$-spin circulation; it also carries a uniform particle circulation when $S_z\neq0$. The out-of-plane spin current is independent of $S_z$ and therefore can remain finite as $S_z\to0$, whereas the particle current then vanishes. The in-plane components are bond-staggered only in $J_x$ via $\eta_{l\to j}$.}

\begin{figure}
    \centering
    \includegraphics[width=\linewidth]{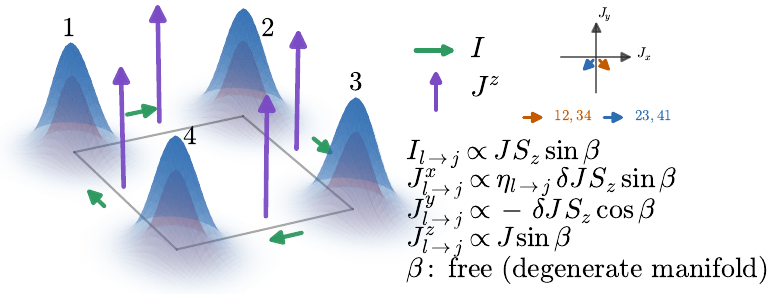}
\caption{\rev{Square-plaquette phase~III shown for one representative of the degenerate manifold (here $\beta=\pi/4$). For $S_z\sin\beta\neq0$, the particle current is uniform and circulates around the plaquette, as shown by the green arrows, and the out-of-plane spin current is also uniform, as shown by the purple arrows. The inset displays the in-plane Stokes-space spin current: the bond pair $(1\!\to\!2,\,3\!\to\!4)$ and the pair $(2\!\to\!3,\,4\!\to\!1)$ have opposite $J_x$ components but the same $J_y$ component. Varying $\beta$ rotates the in-plane pattern and rescales $I$ and $J^z$; at $\sin\beta=0$, the circulating particle and $z$-spin currents vanish without changing the phase label.}}\label{fig:square_phaseIII_3d}
\end{figure}

Changing $\beta$ selects a different representative of the same degenerate Phase~III manifold (see Fig.~\ref{fig:square_phaseIII_3d}): it changes the observable current pattern, with $I$ and $J^z$ scaling as $\sin\beta$ and $\bm{J}^{\perp}$ rotating between its $x$ and $y$ components, but it does not change the phase label.

\paragraph*{Numerical maps and analytic line cuts.}
Figure~\ref{fig:square_heatmaps_three_panel} shows three current maps for the bond $1\!\to\!2$ at fixed $J/Un=-1$, and Fig.~\ref{fig:square_linecuts_currents} shows the corresponding analytic line cuts based on Eqs.~\eqref{eq:square_phaseI_currents}--\eqref{eq:square_phaseIII_currents}. Because Phases~I and III have a free angle $\beta$, numerical minimization may return different representatives of the degenerate manifold at nearby parameter points; this produces the visible speckling in the $\beta$-dependent heat-map observables. The in-plane magnitude $J^{\perp}_{1\to2}$ is insensitive to this continuous $\beta$ choice, and the Phase~II parts of the $J^{\perp}_{1\to2}$ and $J^z_{1\to2}$ maps are shown on one symmetry-related branch. For the analytic line cuts, discontinuities appear only at the physical phase boundaries; in the $\beta$-degenerate phases we either omit observables that are not fixed by the conservative model or show one representative branch explicitly, as for the dotted Phase~III segment of $I_{1\to2}$.

\subsection{Currents in a pentagonal ring}
\label{subsec:curr-pentagon}

For larger regular rings, a full analytic classification of all equilibrium branches becomes impractical. Thus, for $N$-edge polygons with $N\geq 5$ we describe the phases only in terms of the corresponding observables without specifying the exact spin configurations. In addition to the edge currents we evaluate additional metrics. First, we introduce two winding numbers,
\[
W_{\pm}=\frac{1}{2\pi}\sum_{j=1}^{N}\mathrm{Wrap}_{\pi}
\!\left(\varphi_{j+1,\pm}-\varphi_{j,\pm}\right),
\]
where $\varphi_{j,\pm}$ are the component phases from Eq.~\eqref{eq:spinor_param}, $j+1$ is understood modulo $N$, and $\mathrm{Wrap}_{\pi}$ denotes reduction to the principal interval $(-\pi,\pi]$. We also use the branch-invariant common-phase coherence metric:
\[
m_{\Phi}=
\frac{1}{N}\left|
\sum_{j=1}^{N} e^{i2\Phi_j}
\right|.
\]
The winding numbers $(W_+,W_-)$ identify the discrete topological properties of the ground states, while $m_{\Phi}$ resolves the degree of the common-phase coherence. In particular, $m_{\Phi}=1$ corresponds to a perfect locking of the doubled common phase $2\Phi_j=\varphi_{j,+}+\varphi_{j,-}$ across the ring, while $m_{\Phi}=0$ means that these phase factors cancel globally and no net common-phase coherence remains.
The sites are numbered clockwise
$1\!\to\!2\!\to\!3\!\to\!4\!\to\!5\!\to\!1$, and the edge $1$--$5$ is oriented along the $x$ axis.
For simplicity of representation, we show only $W_-$ in the phase maps. 
 When needed, we also monitor the site-to-site variation of $S_{jz}$ to isolate small symmetry-broken regions in the phase-diagram.

\begin{figure}[t]
    \centering
    \includegraphics[width=\linewidth]{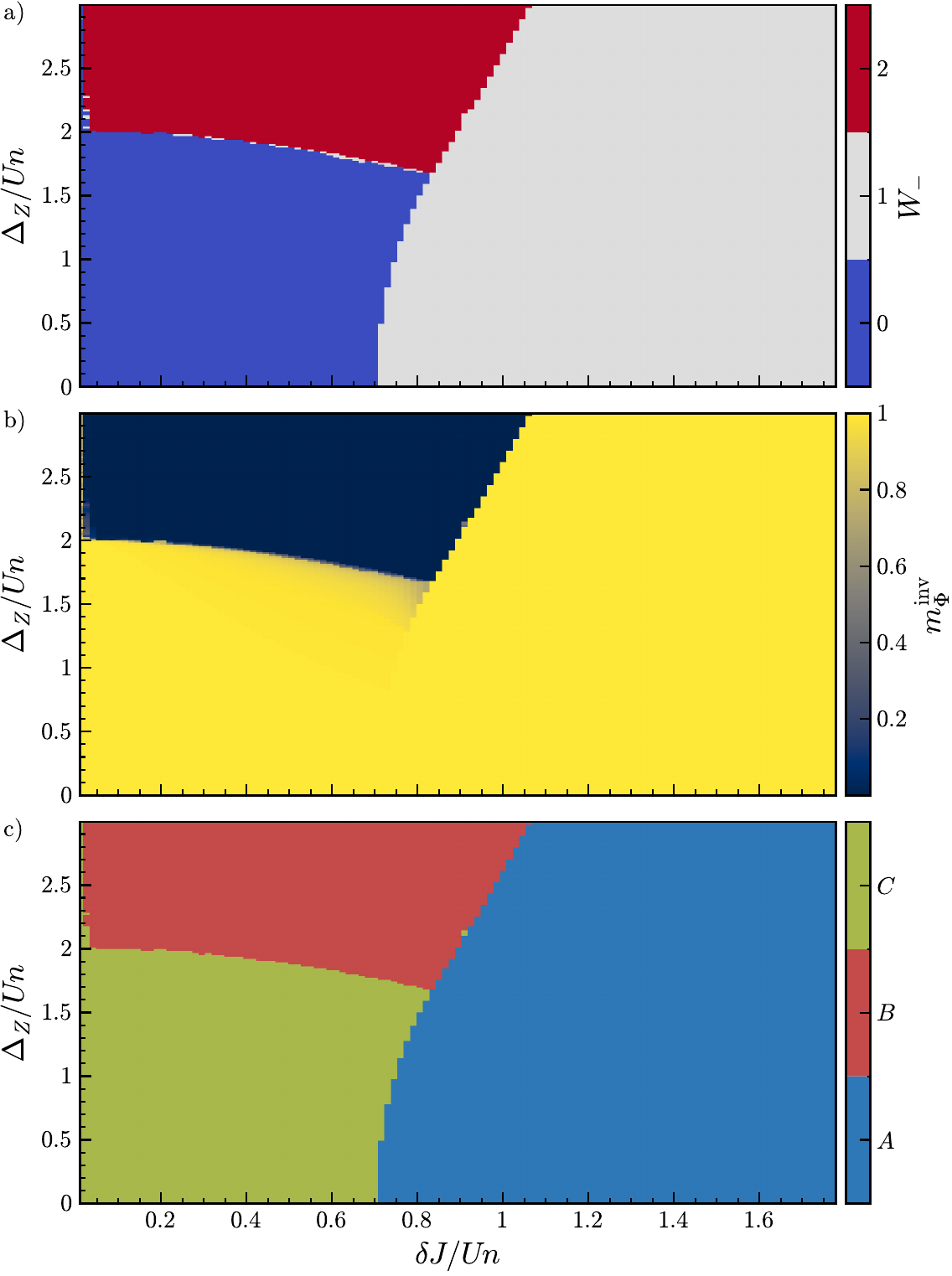}
    \caption{\rev{Metric diagnostics for the pentagonal ring in the same parameter scan as Fig.~\ref{fig:pentagon_heatmaps_three_panel}. Panel~(a) shows the winding number $W_{-}$ of the $\sigma_-$ circular component; panel~(b) shows the branch-invariant common-phase coherence metric $m_{\Phi}$; panel~(c) combines these metrics into the compact domain classification. Domain A is the hidden-vortex-like circulating sector with coherent common phase, domain B is the semi-vortex-like sector with suppressed $m_{\Phi}$ and a large in-plane spin-current magnitude, and domain C is the topologically trivial, bond-textured (asymmetric) sector.}}
    \label{fig:pentagon_phase_diagram_dominant}
\end{figure}

\begin{figure}[t]
\centering
\includegraphics[width=\linewidth]{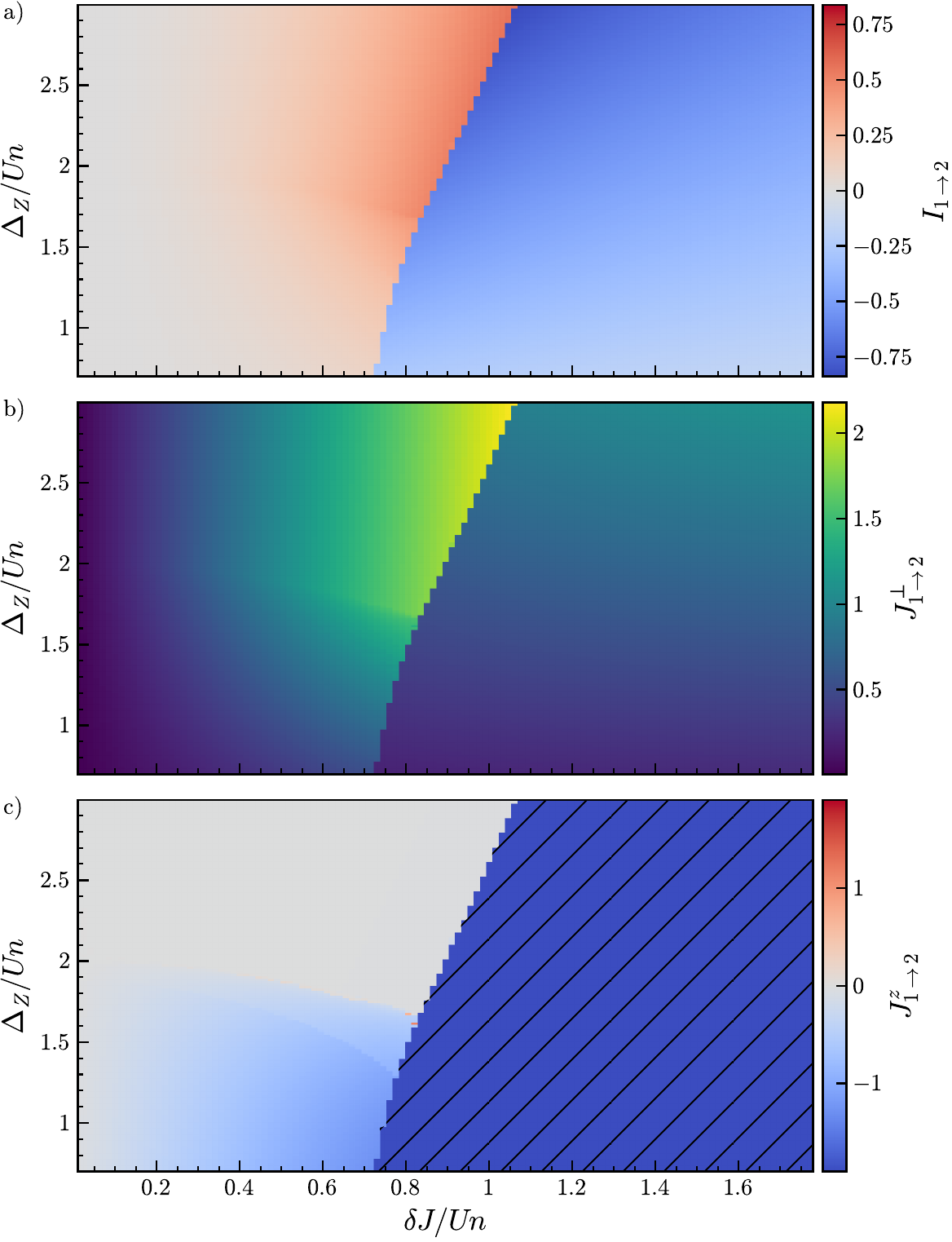}
\caption{\rev{Bond-resolved current maps for a pentagonal ring at fixed $J/Un=-1$, plotted in the parameter plane $(\delta J/Un,\Delta_Z/Un)$ for the directed bond $1\!\to\!2$. All values are currents per particle in the energy-minimizing stationary state: (a) particle current $I_{1\to2}$, (b) in-plane spin-current magnitude $J^{\perp}_{1\to2}$, and (c) out-of-plane spin-current component $J^z_{1\to2}$. The right-hand A domain is identified by a nearly parameter-independent negative $J^z_{1\to2}$ plateau, $J^z_{1\to2}\approx -1.9$, marked by hatching in panel~(c). The upper-left B domain has a large in-plane spin-current magnitude and weak $J^z$, while the lower-left C domain has weak particle flow and a bond-textured $J^z$ response.}}
\label{fig:pentagon_heatmaps_three_panel}
\end{figure}

Figure~\ref{fig:pentagon_phase_diagram_dominant} provides the primary metric-based separation of the pentagon phases. Our metrics distinguish three main domains which we label A, B and C. Domain A is the uniform circulating regime: in the present convention of clockwise site numbering, it is characterized by $(W_+,W_-) = (-1,+1)$ with $m_\Phi\simeq1$, and matches the current pattern of a negative $J^z$ plateau together with, at finite circular polarization, a nearly bond-uniform anticlockwise particle circulation. Domain B belongs to $(W_+,W_-)=(0,+2)$, has strongly suppressed $m_\Phi$, and is identified by a clockwise particle circulation, a large in-plane spin-current response, and nearly vanishing $J^z$, see Fig.~\ref{fig:pentagon_heatmaps_three_panel}. Domain C has $(W_+,W_-)=(0,0)$ and again $m_\Phi\simeq1$, but it differs sharply from A by its weak particle transport and strongly edge-dependent $J^z$. Within this broad trivial sector one may still encounter narrower more anisotropic textures with finite site-to-site polarization nonuniformity, but we do not separate them as an independent phase here.

Note that in the C domain, $J^z_{l\to j}$ is not uniform: its sign and magnitude depend strongly on the edge number, so the ring average remains small even though the local bond currents themselves are appreciable.

In this sense, phase A may be viewed as the larger-polygon analogue of the hidden-vortex branch of the triangle and the circulating state in the square (Phase~III): in the hidden-vortex triangle and in the circulating representatives of the square Phase III manifold ($\sin{\beta}\neq 0$), the loop carries a uniform out-of-plane spin current and, at finite circular polarization, a finite particle circulation.

Phase B is best regarded as the continuation of the triangular semi-vortex regime (Phase~II), since the winding is carried by only one circular component and the transport is dominated by the in-plane spin-current channel while $J^z$ remains weak. By contrast, the trivial sector C does not admit an equally clean counterpart in the triangle or square and is better treated as a genuinely larger-ring weak-current, bond-textured branch.

\paragraph*{Current maps and metric diagnostics.}
Having identified the phases through the metric maps, we now return to the bond currents themselves. Figure~\ref{fig:pentagon_heatmaps_three_panel} shows that the same three domains can be identified in the current maps. At large $\delta J$ (right domain), the representative edge carries an anticlockwise $I_{1\to2}$ and a parameter-independent negative $J^z_{1\to2}$, while $J^{\perp}_{1\to2}$ remains finite but moderate. In the upper-left domain, the particle circulation is clockwise, the in-plane spin-current magnitude is large, while $J^z_{1\to2}$ is nearly suppressed. In the lower-left domain, the particle transport remains weak but is typically clockwise, while $J^z$ becomes strongly edge-dependent, indicating that no simple uniform $z$-spin circulation survives. In Fig.~\ref{fig:pentagon_heatmaps_three_panel}, panels (b,c) therefore use one canonical representative branch for $J^{\perp}_{1\to2}$ and $J^z_{1\to2}$; this branch choice changes only which degenerate representative is displayed, not the phase label or the energy.

\subsection{Currents in a hexagonal ring}
\label{subsec:curr-hexagon}
\begin{figure}[!t]
    \centering
    \includegraphics[width=\linewidth]{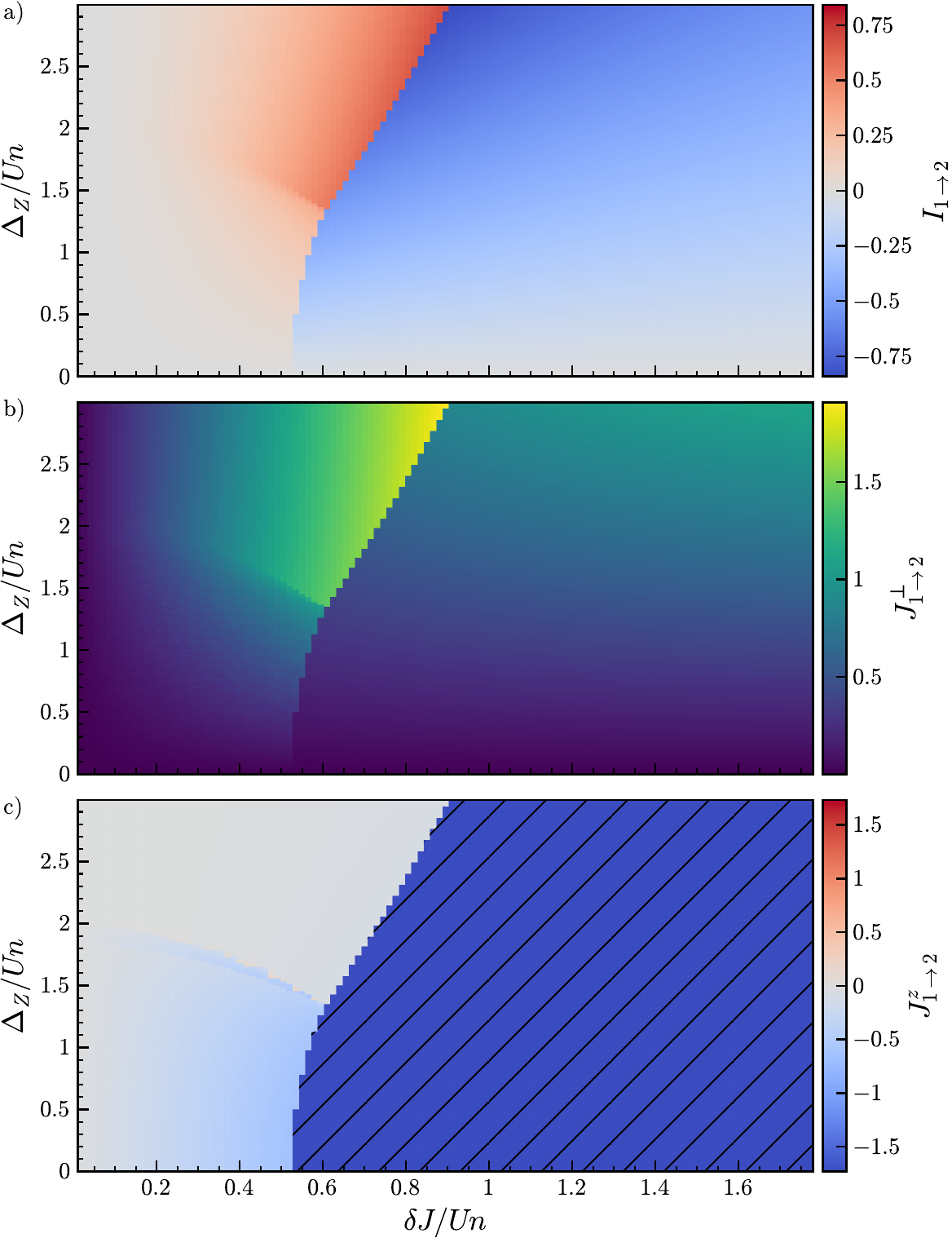}
    \caption{\rev{Bond-resolved current maps for a hexagonal ring at fixed $J/Un=-1$, plotted in the parameter plane $(\delta J/Un,\Delta_Z/Un)$ for the directed bond $1\!\to\!2$. Panels show currents per particle in the energy-minimizing stationary state: (a) $I_{1\to2}$, (b) $J^{\perp}_{1\to2}$, and (c) $J^z_{1\to2}$. The right-hand A domain again forms a nearly parameter-independent negative $J^z_{1\to2}$ plateau, now close to $-\sqrt{3}\approx -1.73$, as expected for a hidden-vortex-like circulation on a six-site ring. The upper-left B domain has a large in-plane spin-current magnitude, while the lower-left C domain is bond textured. In the C-like sector the six cyclically equivalent energy-degenerate orientations are represented by one canonical branch on the selected edge.}}
    \label{fig:hexagon_heatmaps_three_panel}
\end{figure}

\begin{figure}[!t]
    \centering
    \includegraphics[width=\linewidth]{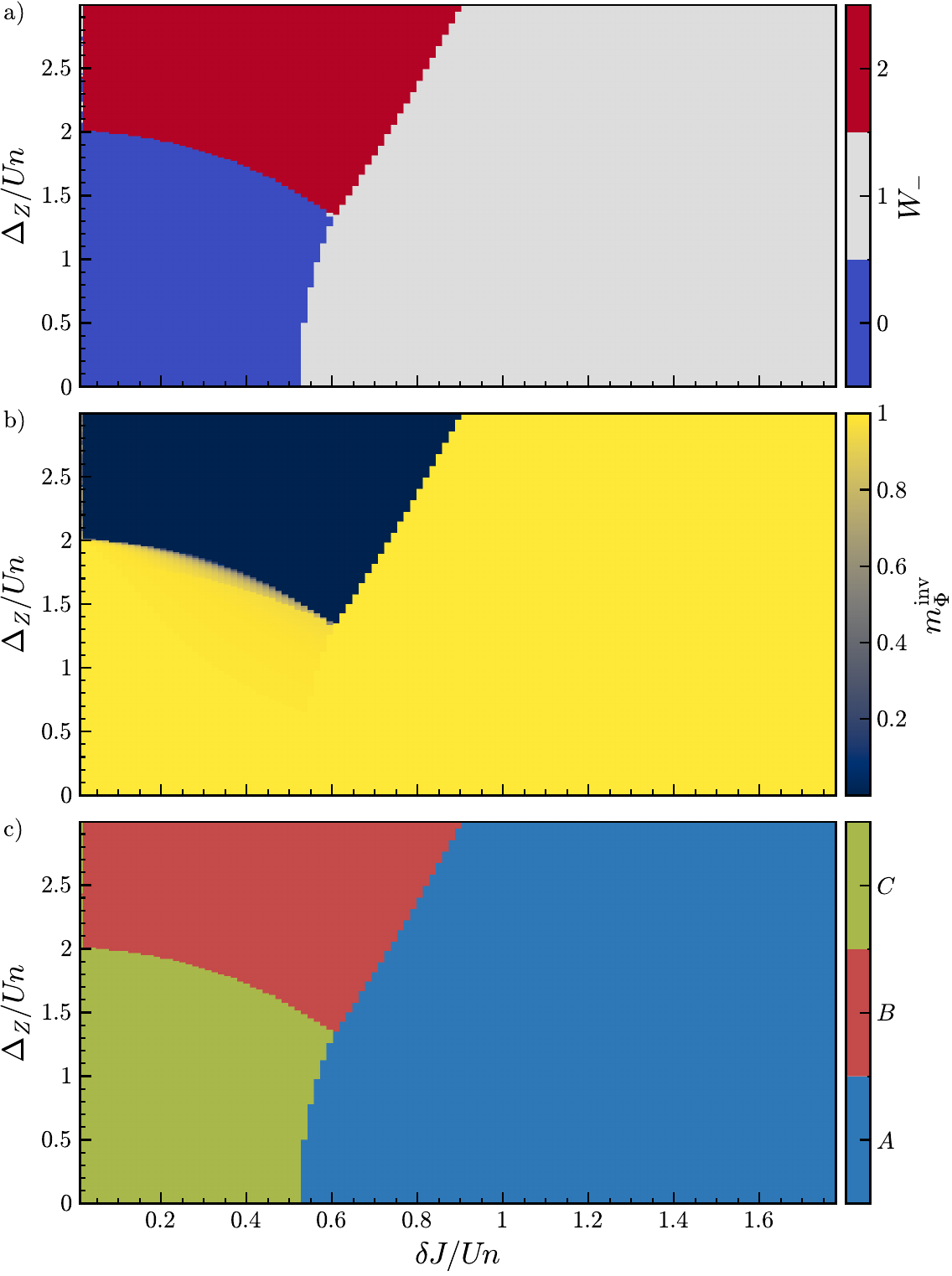}
    \caption{\rev{Metric diagnostics for the hexagonal ring in the same parameter scan as Fig.~\ref{fig:hexagon_heatmaps_three_panel}. Panel~(a) shows the winding number $W_{-}$ of the $\sigma_-$ circular component, panel~(b) shows the branch-invariant common-phase coherence metric $m_\Phi$, and panel~(c) gives the compact A/B/C domain classification. The same interpretation as in the pentagon applies: A is hidden-vortex-like, B is semi-vortex-like with suppressed common-phase coherence, and C is a topologically trivial bond-textured sector. Minor transition pixels and narrow C-like symmetry-broken pockets are absorbed into the broad C domain.}}
    \label{fig:hexagon_metrics_three_panel}
\end{figure}

We now repeat the same analysis for a hexagon network of condensates.
Here, the sites are again numbered clockwise and the edge $1$--$6$ is taken along the $x$-axis.  Figures~\ref{fig:hexagon_heatmaps_three_panel} and \ref{fig:hexagon_metrics_three_panel} reveal the same three macroscopic phases as in the pentagon: the hidden-vortex phase A with $(W_+,W_-) = (-1,+1)$ and $m_\Phi\simeq1$, the semi-vortex state B $(W_+,W_-)=(0,+2)$ with strongly suppressed $m_\Phi$ and a generalized asymmetric phase C with topologically trivial windings and strong phase coherence.
As in the pentagon, the lower-left C-like sector contains cyclically equivalent energy-degenerate orientations of a bond-textured state. A raw fixed-edge scan can jump between these representatives; therefore, in Fig.~\ref{fig:hexagon_heatmaps_three_panel}, panels (b,c) show one canonical representative branch for $J^{\perp}_{1\to2}$ and $J^z_{1\to2}$. This branch choice affects only which degenerate representative is displayed on the selected edge, not the phase label or the energy.

\rev{The current and metric maps reveal a systematic shift of the phase boundaries as the number of edges increases. In particular, the critical spin-flip strength associated with the onset of the hidden-vortex-like phase A decreases with $N$. In the next section we discuss how this trend continues in larger polygonal polariton networks.}

\subsection{Toward the continuum-ring limit}
\label{subsec:continuum-ring}

The phase diagrams of the pentagon, hexagon, heptagon, octagon, and nonagon suggest that the current-based phase structure of regular condensate rings belongs to a common finite-$N$ sequence approaching a continuum-ring regime. We do not discuss the heptagon, octagon, and nonagon as separate case studies, because their phase portraits remain qualitatively similar to those of $N=5$ and $6$; the main systematic change is a horizontal drift of the dominant phase boundaries in the raw parameter plane.

Across $N=5,6,7,8,9$ the dominant stationary phases admit a common current-based interpretation. Phase A is a chiral circulation branch with strong bond-uniform $J^z$ and, at finite circular polarization, a nearly bond-uniform particle current; depending on the adopted clockwise/counterclockwise labeling convention, its winding sector is written as $(W_+,W_-)=(+1,-1)$ or $(-1,+1)$, with the directed-current signs changing accordingly. Under the branch-invariant scalar metric used here, this right-hand circulating sector remains a single coherent macro-domain across the whole sequence. Phase B carries a strong nearly uniform in-plane spin-current response and suppressed $J^z$; correspondingly, its winding sector is written as $(W_+,W_-)=(0,-2)$ or $(0,+2)$, again depending on the orientation convention. Phase C is characterized by $(W_+,W_-)=(0,0)$ and has a weak particle transport together with strongly bond-textured $J^z$. More strongly symmetry-broken textures may still appear inside this broad C sector, but we do not treat them as separate macroscopic phases.

A direct comparison of the raw phase diagrams shows that the main boundaries shift systematically toward lower $\delta J/Un$ as $N$ increases. This is not an effect of the polygon orientation, since each site still has two nearest neighbours on the ring; rather, it reflects the decrease of variation in the angle between adjacent sites with the increase of $N$. Because the TE--TM coupling enters through a double-angle structure, the relevant discrete angular scale is controlled by $4\pi/N$ rather than by the scalar-ring angle $2\pi/N$. Therefore, we introduce the effective rescaled spin-flip strength
\begin{equation}
\widetilde{\delta J}
=
\frac{\delta J}{\bigl[1-\cos(4\pi/N)\bigr]Un},
\label{eq:deltaJ_rescaled}
\end{equation}
where the factor in the denominator is motivated by the bond-energy contribution of the TE--TM-splitting energy on a polygon network. On a regular $N$-edge polygon, neighbouring bonds differ by the doubled geometric angle $\Delta\theta=4\pi/N$. Since the TE--TM coupling enters through trigonometric factors of this bond angle, the quantity $1-\cos(\Delta\theta)$ simply measures how different two neighbouring bond orientations are: it is zero when the orientations coincide and increases as the angular mismatch grows. Thus the relevant finite-$N$ geometric scale is $1-\cos(4\pi/N)$.
For large $N$ one has
\begin{equation}
1-\cos(4\pi/N) = \frac{8\pi^2}{N^2} + O(N^{-4}),
\end{equation}
so that Eq.~\eqref{eq:deltaJ_rescaled} is asymptotically equivalent to an $N^2$ rescaling,
\begin{equation}
\widetilde{\delta J}
\sim
\frac{N^2}{8\pi^2}\,\frac{\delta J}{Un}.
\end{equation}
Equivalently, the position of the phase boundary separating the A phase should be shifted towards lower values of $\delta J$  approximately as $(\delta J/Un)_c\propto N^{-2}$.
\begin{figure}
    \centering
    \includegraphics[width=\linewidth]{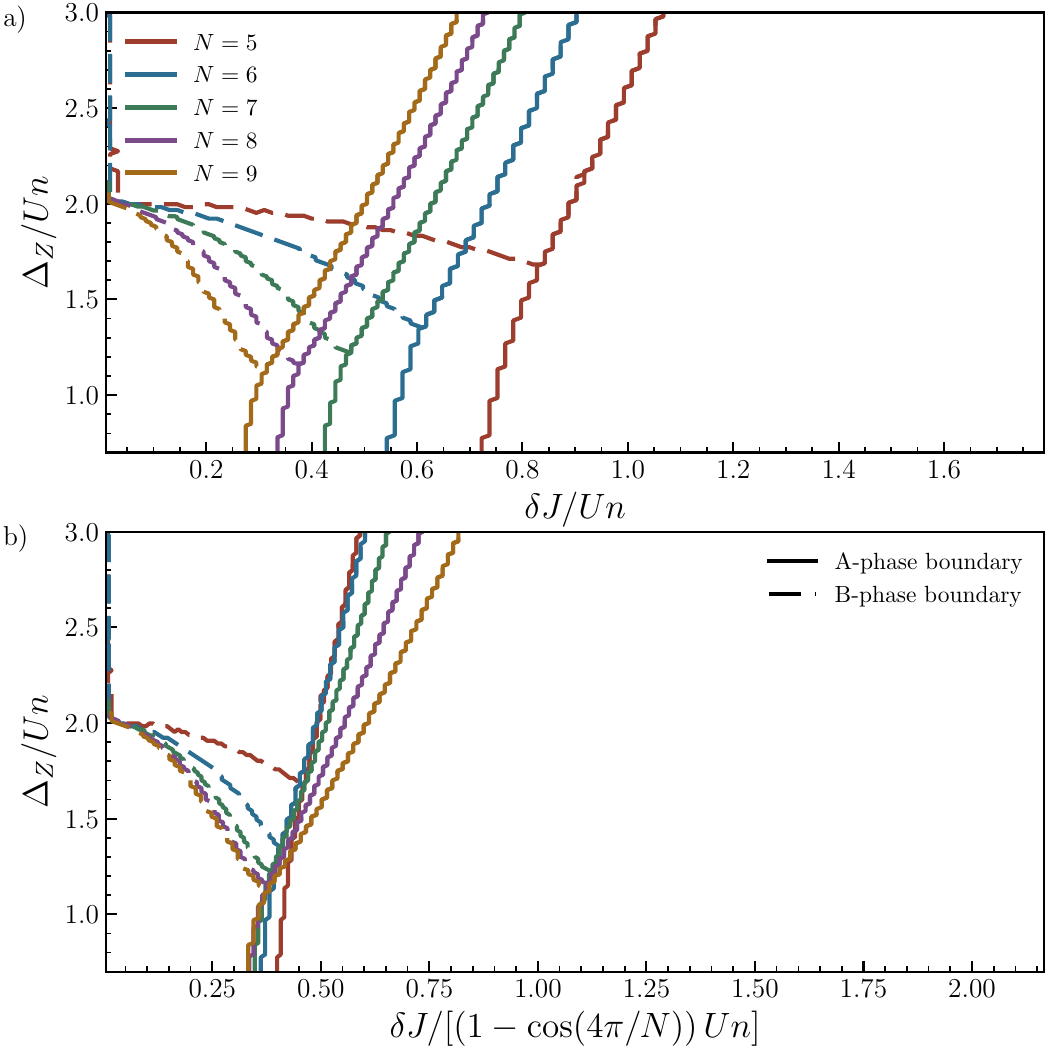}
\caption{\rev{Dominant phase-boundary overlay for the pentagon ($N=5$), hexagon ($N=6$), heptagon ($N=7$), octagon ($N=8$), and nonagon ($N=9$). Panel~(a) shows the boundaries in the raw parameter plane $(\delta J/Un,\Delta_Z/Un)$; panel~(b) shows the same data after rescaling the horizontal axis according to Eq.~\eqref{eq:deltaJ_rescaled}. Colors distinguish $N$, while line styles distinguish the extracted boundaries: solid curves mark the onset of the A domain and dashed curves mark the B-domain boundary. The rescaling removes most of the systematic horizontal drift of the A boundary, but the high-field part still shows finite-$N$ fan-out; the B boundary aligns only partially and should be interpreted as a qualitative trend rather than an exact collapse.}}
\label{fig:continuum_overlay_scaling}
\end{figure}

For example, keeping $\widetilde{\delta J}$ fixed implies that the raw coupling should be transformed between neighbouring polygons as
\begin{equation}
\left(\frac{\delta J}{Un}\right)_{N+1}
=
\left(\frac{\delta J}{Un}\right)_{N}
\frac{1-\cos\bigl(4\pi/(N+1)\bigr)}{1-\cos\bigl(4\pi/N\bigr)}.
\end{equation}
Thus, if one chooses $(\delta J/Un)_{N=5}=1$, the corresponding values that keep the same rescaled coordinate are $(\delta J/Un)_{N=6}\approx 0.82918$, $(\delta J/Un)_{N=7}\approx 0.67579$, 
$(\delta J/Un)_{N=8}\approx 0.55279$.
In this sense, the quantity that remains approximately invariant across $N$ is not the raw $\delta J/Un$ itself, but the rescaled combination $\widetilde{\delta J}$.

Among the simple rescalings we tested, Eq.~\eqref{eq:deltaJ_rescaled} gives the best one-parameter collapse of the dominant A-phase onset. For $N=5,6,7,8,9$ the relative RMS spread of this boundary is reduced from about $0.242$ in the raw variables to about $0.085$ after rescaling. The improvement is not perfectly uniform along the full contour: the low- and intermediate-field part of the A boundary aligns well already for $N=7,8,9$, whereas its upper high-$\Delta_Z/Un$ segment still shows a noticeable residual fan-out. The plotted B boundary also shows partial alignment, but different sides of the B domain collapse with different quality. Thus the rescaling captures the leading horizontal drift of the phase portrait without making every boundary segment exactly $N$-independent.

The effect of the rescaling is shown in Fig.~\ref{fig:continuum_overlay_scaling}. In the raw variables, the onset of the right-hand A phase drifts visibly with $N$. After rescaling, these dominant boundaries become much more closely aligned overall, while the remaining discrepancies are consistent with subleading finite-$N$ effects and the residual high-field fan-out of the upper A segment. We therefore interpret the observed collapse not as a strict proof of the $N\to\infty$ limit, but as numerical evidence for a common finite-$N$ trend consistent with a continuum-like phase structure.

The finite polygons therefore show three recurring current regimes. Phase A is a chiral branch with strong $z$-spin flow and, at finite circular polarization, finite particle current. Phase B has a large in-plane spin-current response and suppressed $J^z$. Phase C is a weak-current branch with a bond-textured out-of-plane spin-current response. The smaller symmetry-broken windows inside C are best viewed as finite-$N$ refinements rather than as separate continuum phases.

\section{Conclusion}\label{sec:Conclusion}

We have described equilibrium states in tunnel-coupled spinor polariton rings through their bond-resolved particle and spin currents. Starting from the conservative spinor Hamiltonian, we separated the spin current into transport and interfacial torque contributions and used this decomposition to interpret the stationary states obtained by energy minimization.

For the minimal geometries with known analytical phase structure, we derived explicit current textures. The dyad supports no particle current but may carry a purely in-plane spin current. In the triangular cell, the asymmetric, semi-vortex, and hidden-vortex states acquire clearly distinct particle- and spin-current signatures. In the square plaquette, the three equilibrium phases demonstrate three equally distinct transport regimes: bond-staggered torque-induced spin current without particle flow, bond-asymmetric spin transport with vanishing particle current, and a free-angle manifold containing representatives with uniform $z$-spin currents and, at finite circular polarization, uniform particle currents.

\rev{For larger rings, where a full analytic phase-by-phase treatment is no longer practical, we showed explicitly for $N=5$ and $N=6$ (and documented $N=7$--$9$ in the Appendix) that bond currents remain useful phase diagnostics when combined with scalar metrics such as winding numbers and branch-invariant common-phase coherence.} The current maps and the metric maps identify the same macroscopic phases and provide a compact classification in terms of particle circulation, in-plane spin-current response, and out-of-plane spin-current texture.

Finally, comparison of the pentagon, hexagon, heptagon, octagon, and nonagon reveals a systematic finite-$N$ trend toward a common continuum-like phase portrait.
This convergence emerges after the rescaling of Eq.~\eqref{eq:deltaJ_rescaled}, which reflects the double-angle structure of the TE--TM coupling on a ring. Under this scaling, the dominant branches become substantially more aligned, although residual finite-$N$ discrepancies remain, most visibly in the upper high-field part of the A-domain boundary. This indicates that the discrete polygons organize into leading current phases consistent with the finite-$N$ scaling toward a continuum-like ring structure.

These current patterns give experimentally accessible signatures of the equilibrium states in polariton graphs. They distinguish persistent particle circulation, hidden spin-current responses, torque-assisted spin conversion, and bond-dependent spin-current patterns. The next question is which of these conservative patterns remain stable in a driven-dissipative setting.


\begin{acknowledgments}
We thank P.G. Lagoudakis for valuable discussion. A.N.O.\ and A.V.Y.\ acknowledge support from the Russian Science Foundation.
A.K.\ was supported by the Icelandic Research Fund (Grant No.~2410550).
\end{acknowledgments}

\appendix
\section{Supplementary Maps for Larger Polygons}\label{app:larger-polygons}

\begin{figure}
    \centering
    \includegraphics[width=\linewidth]{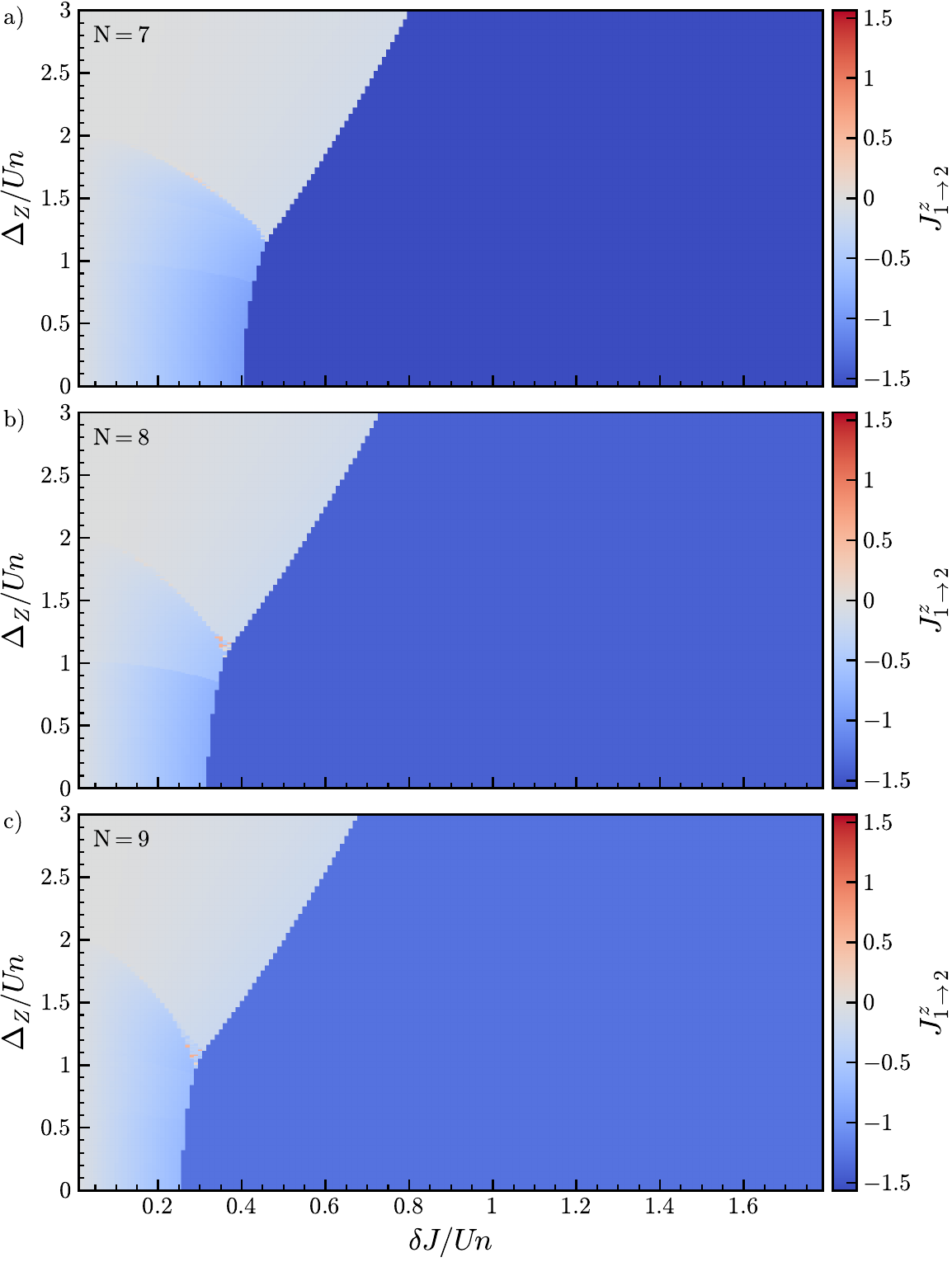}
    \caption{\rev{Out-of-plane spin-current maps $J^z_{1\to2}$ for the heptagon ($N=7$), octagon ($N=8$), and nonagon ($N=9$) at fixed $J/Un=-1$, plotted in the same parameter plane and clockwise numbering convention as the main-text polygon maps. The right-hand A domain retains a broad negative $J^z_{1\to2}$ plateau, the upper-left B domain keeps $J^z_{1\to2}$ strongly suppressed, and the lower-left C domain shows a bond-textured response. In the C-like sector, one canonical representative branch is shown for the selected edge $1\!\to\!2$, as in the pentagon and hexagon maps.}}
    \label{fig:appendix_jz_789}
\end{figure}

\begin{figure}
    \centering
    \includegraphics[width=\linewidth]{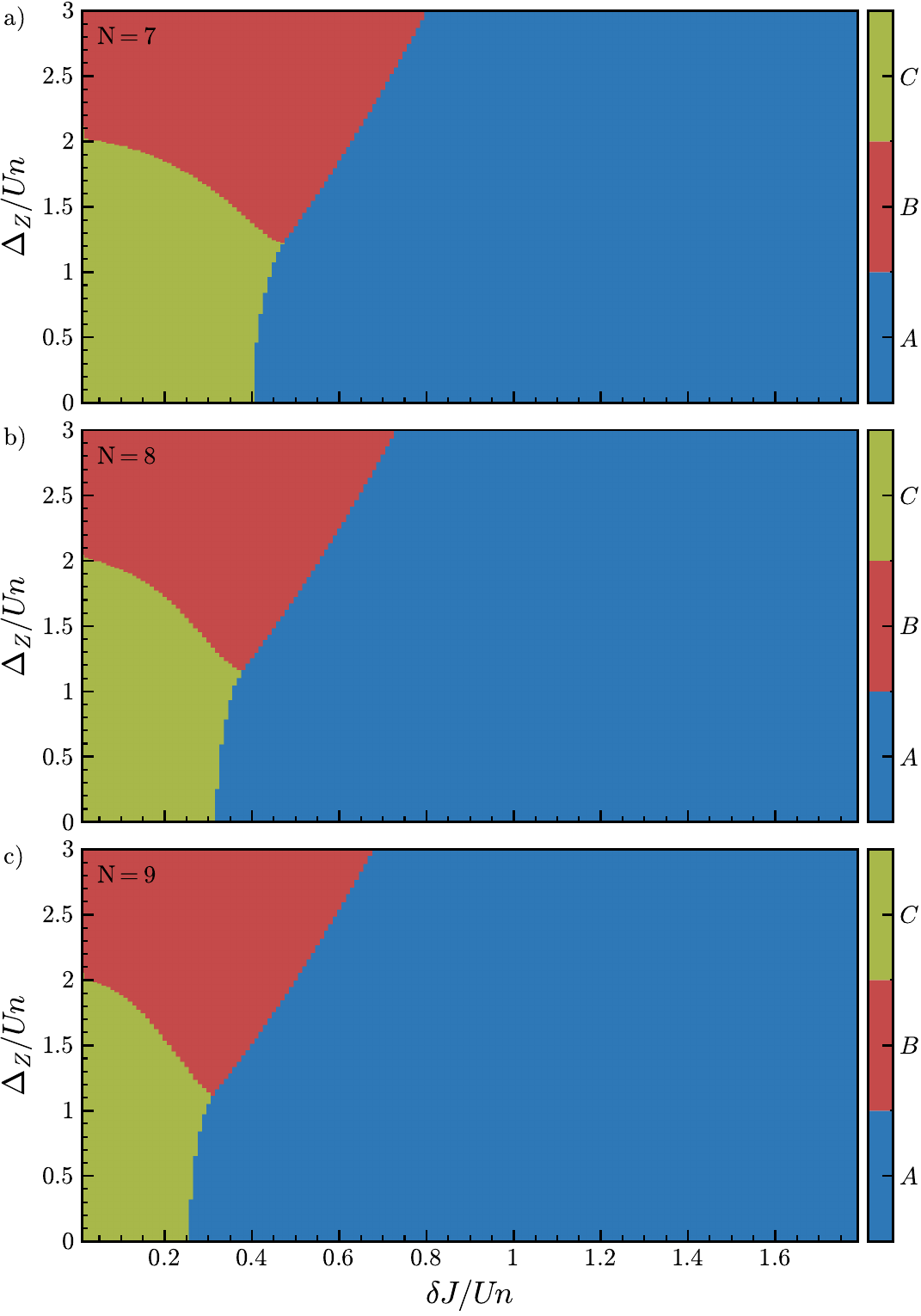}
    \caption{\rev{Compact dominant phase diagrams for the heptagon ($N=7$), octagon ($N=8$), and nonagon ($N=9$), shown in the same parameter plane as Fig.~\ref{fig:appendix_jz_789}. Only the three macroscopic domains are retained: A denotes the hidden-vortex-like circulating sector, B the semi-vortex-like sector with suppressed common-phase coherence, and C the topologically trivial bond-textured sector. Narrower C-like symmetry-broken pockets are absorbed into the broader C domain to match the compact classification used in the main text.}}
    \label{fig:appendix_phase_789}
\end{figure}

For completeness, we collect here the corresponding maps for the heptagon ($N=7$), octagon ($N=8$), and nonagon ($N=9$), evaluated in the same clockwise convention and in the same parameter plane as in the main text. These supplementary panels are intended to document the smooth continuation of the larger-ring sequence beyond $N=6$ without interrupting the main discussion by three additional case studies.

Figure~\ref{fig:appendix_jz_789} shows that the bond-resolved out-of-plane spin current follows the same qualitative organization already established for the pentagon and hexagon. The right-hand phase A retains a broad negative $J^z_{1\to2}$ plateau, the upper-left phase B keeps $J^z_{1\to2}$ strongly suppressed, and the lower-left phase C develops the bond-textured response discussed in the main text. The principal finite-$N$ effect is therefore not the appearance of new macroscopic domains, but the gradual horizontal drift and reshaping of the same three transport sectors as $N$ increases.

The phase maps in Fig.~\ref{fig:appendix_phase_789} make this continuation explicit. No additional robust macroscopic phase emerges between $N=7$ and $9$: the same A, B, and C domains persist, while their boundaries move systematically toward lower raw values of $\delta J/Un$. This is the same finite-$N$ trend summarized in the rescaled overlay of Fig.~\ref{fig:continuum_overlay_scaling}, now displayed here in the underlying bond-current and dominant-domain representations for each larger polygon separately.

\clearpage
\bibliographystyle{apsrev4-2}
\bibliography{references}

\end{document}